\documentclass[aps,prx,twocolumn,english,showpacs,superscriptaddress,amssymb,amsfonts]{revtex4}
\usepackage[T1]{fontenc}
\usepackage[latin9]{inputenc}
\usepackage{amsmath}
\usepackage{dsfont}
\usepackage{amstext}

\usepackage{tocvsec2}

\usepackage{amssymb}
\usepackage{amsbsy}
\usepackage{amsthm}
\usepackage{epsfig}
\usepackage{graphicx}
\usepackage{bbm}
\usepackage{hyperref}
\usepackage{color}
\usepackage{multirow}
\usepackage{pdfsync}
\def\beq{\begin{equation}}
\def\eeq{\end{equation}}

\usepackage{mdframed}
\usepackage{lipsum}
\newmdtheoremenv{theorem}{Conjecture}

\newcommand{\bsphi}{{\boldsymbol{\phi}}}

\newcommand{\bst}{{\mathcal{T}}}

\newcommand{\mcg}{{\mathcal{G}}}

\newcommand{\bsq}{{\boldsymbol{q}}}

\newcommand{\ie}{{\emph{i.e.~}}}
\makeatletter

\newcommand{\Rmnum}[1]{\expandafter\@slowromancap\romannumeral #1@}
\makeatother
\newcommand{\imth}{\hspace{1pt}\mathrm{i}\hspace{1pt}}

\newcommand{\eg}{{\emph{e.g.~}}}

\newcommand{\mbz}{{\mathbb{Z}}}
\newcommand{\bea}{\begin{eqnarray}}
\newcommand{\eea}{\end{eqnarray}}
\newcommand{\bpm}{\begin{pmatrix}}
\newcommand{\epm}{\end{pmatrix}}
\newcommand{\bal}{\begin{aligned}}
\newcommand{\eal}{\end{aligned}}

\newcommand{\dket}[1]{|{#1}\rangle}

\newcommand{\mcp}[1]{\mathcal{P}_{#1}}
\newcommand{\gspt}[1]{|{#1}\rangle_{\mathcal{G}_y}}

\makeatother

\begin{document}
\title{Classification and surface anomaly of glide symmetry protected topological phases in three dimensions}

\author{Fuyan Lu}
\author{Bowen Shi}
\author{Yuan-Ming Lu}
\affiliation{Department of Physics, The Ohio State University, Columbus, OH 43210, USA}

\date{\today}

\begin{abstract}
We study glide protected topological (GSPT) phases of interacting bosons and fermions in three spatial dimensions with certain on-site symmetries. They are crystalline SPT phases, which are distinguished from a trivial product state only in the presence of non-symmorphic glide symmetry. We classify these GSPT phases with various on-site symmetries such as $U(1)$ and time reversal, and show that they can all be understood by stacking and coupling two-dimensional short-range-entangled phases in a glide-invariant way. Using such a coupled layer construction we study the anomalous surface topological orders of these GSPT phases, which gap out the two-dimensional surface states without breaking any symmetries. While this framework can be applied to any non-symmorphic SPT phase, we demonstrate it in many examples of GSPT phases including the non-symmorphic topological insulator with ``hourglass fermion'' surface states.
\end{abstract}

\pacs{}

\maketitle
\tableofcontents



\section{Introduction}

After the discovery of topological insulators (TIs) with surface Dirac fermions\cite{Hasan2010,Hasan2011,Qi2011}, a large class of symmetry-protected topological (SPT) phases\cite{Chen2012b,Senthil2015} have been revealed to exist beyond Landau's paradigm. Despite the absence of anyons and topological orders in the bulk, these SPT phases cannot be continuously connected to a trivial product state as long as certain symmetries are present. One most prominent feature of these SPT phases are the existence of anomalous surface states, that cannot be realized by any symmetric local Hamiltonian in lower spatial dimensions. Take the two-dimensional (2d) surface of a three-dimensional (3d) SPT phase for example, though usually gapless in a weakly-interacting fermion system like TIs, strong interactions can fully gap out the surface states in a symmetric way by developing intrinsic topological orders on the 2d surface\cite{Vishwanath2013,Fidkowski2013,Wang2013a,Metlitski2013,Bonderson2013,Wang2013,Chen2014a,Cho2014}. These gapped surface topological orders however are anomalous, in the sense that anyons carry certain symmetry quantum numbers that are not allowed in a pure 2d state\cite{Chen2016}. Due to bulk-boundary correspondence, both the gapless and gapped 2d surface states serve as diagnosis of the 3d SPT phase.

While SPT phases protected by global (``on-site'') symmetries have been classified and extensively studied\cite{Chen2013,Senthil2015}, ``weak'' interacting SPT phases protected by spatial symmetries are theoretically less understood\cite{Ando2015,Isobe2015,Qi2015a,Song2016,Thorngren2016}. Meanwhile since crystalline symmetries are quite ubiquitous in solids, understanding crystalline SPT phases is important for experimental realization of SPT phases in solid state materials. In this work we focus on crystalline SPT phases protected by non-symmorphic glide symmetry and other global symmetries, such as $U(1)$ charge/spin symmetry and time reversal symmetry $\bst$. In particular, these ``weak'' SPT phases are distinguished from a trivial product state \emph{only} in the presence of glide symmetry, hence coined glide symmetry protected (GSPT) phases.

Previously GSPT phases have been mostly studied in the context of weakly-interacting electrons\cite{Fang2015,Varjas2015,Shiozaki2016,Alexandradinata2016,Wang2016b,Ezawa2016,Kruthoff2016}. In particular the gapless surface states of a 3d glide-protected non-symmorphic TI, coined ``hourglass fermions''\cite{Wang2016b}, has been observed in KHgSb using angle-resolved photoemission spectroscopy\cite{Qian2016}. How do strong electronic correlations modify the classification and physical properties of these non-symmorphic TIs? Here we answer this question by considering a generic interacting system of bosons and/or fermions. Inspired by the general ideas introduced in Ref.\cite{Song2016}, we systematically classify interacting GSPT phases with different on-site symmetries, primarily focusing on $U(1)$ charge/spin and time reversal symmetry. Moreover, we explicitly construct these GSPT phases by coupling an array of 2d SPT layers\cite{Mross2016,Ezawa2016,Fulga2016,Song2016} preserving the global symmetries, in a glide-invariant fashion.

What can strong interactions do to the gapless surface states of a glide-protected TI? The coupled-layer construction provides a simple platform to study interaction effects on the glide-preserving side surface (see \eg FIG. \ref{fig:glide4}), which hosts the gapless surface states. In this work we show that by depositing an array of quantum wires on the side surface, the gapless surface states (including the hourglass fermions as one example) can be fully gapped out, resulting in a surface topological order that preserves glide and all global symmetries. In particular these surface topological orders are anomalous, in the sense that glide symmetry acts on anyons in a way that's not possible in a pure 2d system\cite{Chen2016,Hermele2016}. We explicitly construct these surface topological orders (STOs) by writing down the desired interactions, and discuss why the glide symmetry operation is anomalous in these STOs. The results are summarized in TABLE \ref{tab:a} for bosons and TABLE \ref{tab:b} for fermions.

This article is organized as follows. In section \ref{sec:general strategy} we establish the fixed-point wavefunction of a generic interacting GSPT phase, which naturally leads to the coupled layer construction of an arbitrary GSPT phase. This allows us to classify GSPT phases with any on-site symmetry group $G_0$, and construct their anomalous STOs using the coupled wire construction. We then apply this formulation to 3d GSPT phases of interacting bosons (section \ref{sec:boson}, TABLE \ref{tab:a}) and fermions (section \ref{sec:F}, TABLE \ref{tab:b}). Finally we conclude in section \ref{sec:conclude} by conjecturing a simple but powerful relation (Conj. \ref{thm:3d GSPT<->2d SRE}) between the classification of $(d+1)$-dimensional GSPT phases and classification of $d$-dimensional short-range-entangled phases.

\section{General strategy}\label{sec:general strategy}

While systematic classifications of on-site symmetry ($G_0$) protected topological phases of interacting bosons have been obtained from group cohomology\cite{Chen2013} and cobordism\cite{Kapustin2014,Kapustin2015}, there is no general classification of spatial symmetry protected phases especially for interacting fermions. How to tackle such a complicated problem?

In this section, we establish the ``fixed-point wavefunctions'' of GSPT phases, which are described by an array of decoupled 2d SPT layers arranged in a glide-symmetric way (see FIG. \ref{fig:glide1}). This provides the theoretical foundation for the coupled layer construction for GSPT phases, which will be used extensively to study anomalous surface topological orders later. It also provides a dimensional reduction scheme, which relates the classification of 3d GSPT phases to 2d SPT phases. Below we outline the dimensional reduction argument, generalizing the ideas proposed in Ref.\cite{Song2016}. One major difference is that while Ref.\cite{Song2016} discussed point group symmetry where certain points/lines/planes in space are invariant under all point group operations, there are no invariant points for non-symmorphic spatial symmetries, such as glide considered here.

\subsection{Fixed-point wavefunctions and coupled layer construction of GSPT phases}\label{sec:loc unitary}

\begin{figure}[ht]
\begin{center}
\includegraphics[width=6.5cm]{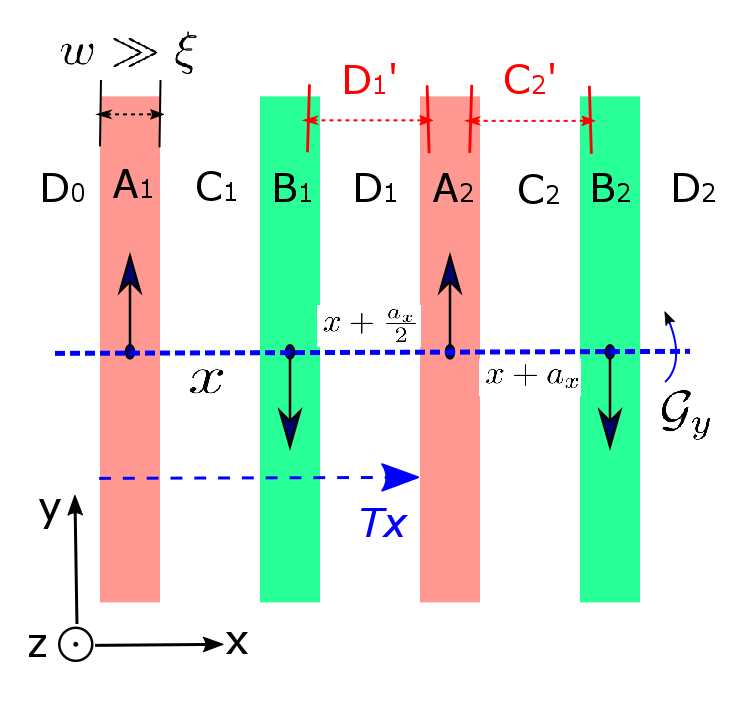}
\end{center}
\caption{The fixed-point wavefunction and coupled layer construction of 3d GSPT phases. The glide operation $\mcg_y$ features a $[010]$ reflection plane and half translation along $\hat x$ direction, satisfying $(\mcg_y)^2=T_x$. The fixed-point wavefunction (\ref{fixed point wf}) is the tensor product of 2d SRE phases in $A_n$ and $B_n$ regions and trivial product state in all other regions.}
\label{fig:glide1}
\end{figure}

We start from a generic gapped system with the full symmetry group
\bea
G_s=\mbz^{\mcg_y}\times G_0=\{(\mcg_y)^n|n\in\mbz\}\times G_0
\eea
generated by global (on-site) symmetry $G_0$ and non-symmorphic glide symmetry $\mcg_y=\{M_{[010]}|(\frac12,0,0)\}$
\bea
(x,y,z)\overset{\mcg_y}\longrightarrow (x+\frac{a_x}2,-y,z).
\eea
as illustrated in FIG. \ref{fig:glide1}, where the arrows point to $\pm\hat y$ directions and $a_x$ represents the length of Bravais lattice primitive vector along $\hat x$ direction. The orientation-reversing glide operation combines mirror reflection w.r.t. [010] plane and translation along $\hat x$ direction by half a Bravais lattice vector. Two glide operation leads to a Bravais translation $T_x=(\mcg_y)^2$ along $\hat x$ direction.

As mentioned earlier, GSPT phases are ``weak'' crystalline SPT states, which can be adiabatically connected to trivial product states if glide symmetry is absent. They are different from the ``strong'' SPT phases, which cannot be adiabatically connected to a product state irrespective of any crystal symmetries, as long as certain on-site/global symmetries are preserved. Mathematically, this means there exists a finite depth quantum circuit $U^{loc}$ that preserves all global symmetries such that
\bea
U^{loc}\dket{\psi}=\dket{0}
\eea
where $\dket{\psi}$ represents any GSPT state and $\dket{0}$ is the trivial product state. Hereafter we assume a finite correlation length $\xi\ll a_x$ in GSPT state $\dket{\psi}$. This assumption can always be satisfied by choosing a large primitive unit cell along $\hat x$ direction, and we expect the topological classification of long-wavelength (infrared) physics to be independent of these microscopic (ultraviolet) details. We divide the physical Hilbert space of this GSPT state into 4 types of regions: $\{A_m,B_m,C_m,D_m\}$ as shown in FIG. \ref{fig:glide1}. In particular these regions are related pairwise by the glide operation:
\bea
&A_m\overset{\mcg_y}\longrightarrow B_m\overset{\mcg_y}\longrightarrow A_{m+1},\\
&C_m\overset{\mcg_y}\longrightarrow D_m\overset{\mcg_y}\longrightarrow C_{m+1}.\notag
\eea
where $m\in\mbz$ is the unit cell index along $\hat x$ direction. In particular we choose the width of $A_n$ and $B_n$ regions to be $w\gg\xi$\footnote{In fact we require the width of each region $\{A_n,B_n,C_n,D_n\}$ to be much larger than the correlation length $\xi$.}, as illustrated in FIG. \ref{fig:glide1}. Keeping a finite value of width $w$ in the thermodynamic limit allows us to treat each $A_n$ (or $B_n$) region as a 2d system extensive along $\hat y$ and $\hat z$ directions.

As shown in FIG. \ref{fig:glide1}, for a finite depth quantum circuit $U^{loc}$, we can always find regions $C_n^\prime\supset C_n$ and $D_n^\prime\supset D_n$, such that
\bea
U^{loc}_{C_n^\prime}\dket{\psi}=\dket{0}_{C_n}\otimes\dket{\psi_{\bar{C}_n}},~~~U^{loc}_{C_n^\prime}\equiv \mcp{C_n^\prime}U^{loc}\mcp{C_n^\prime}
\eea
where $\mcp{R}$ is the projection operator into the Hilbert space of region $R$, and $\bar R$ represents the region outside $R$. In other words, we can always identify local finite depth quantum circuits $U^{loc}_{C_n^\prime}$ that trivializes the a region $C_n$, which locally transform the GSPT state into a trivial product state $\dket{0}_{C_n}$ within region $C_n$. Notice that all these regions do not overlap with each other:
\bea
&C_m^\prime\cap C_n^\prime=D_m^\prime\cap D_n^\prime=0,~~~\forall~m\neq n,\\
&\notag C_m^\prime\cap D_n^\prime=0,~~~\forall~m,n.
\eea
Therefore we can construct the following symmetric finite depth quantum circuit that preserves glide and all on-site symmetries:
\bea\label{sym f.d.q.c.}
&U^{loc}_{CD}\equiv\prod_n U^{loc}_{C_n^\prime}\cdot U^{loc}_{D_n^\prime},\\
&\notag U^{loc}_{D_n^\prime}\equiv\mcg_y U^{loc}_{C_n^\prime}\mcg_y^{-1},~~~U^{loc}_{C_{n+1}^\prime}\equiv\mcg_y U^{loc}_{D_n^\prime}\mcg_y^{-1}.
\eea
such that it trivializes all regions $\{C_n,D_n\}$ for GSPT state $\dket{\psi}$:
\bea\label{fixed point wf}
\dket{\psi_{f.p.}}=U^{loc}_{CD}\dket{\psi}=\prod_n\dket{0}_{C_n\cup D_n}\otimes\dket{\psi_{A_n}}\otimes\dket{\psi_{B_n}}.
\eea
Clearly the above ``fixed-point'' state $\dket{\psi_{f.p.}}$ for a GSPT phase satisfies all symmetries since
\bea\label{coupled layer:glide sym}
\dket{\psi_{B_n}}=\mcg_y\dket{\psi_{A_n}},~~~\dket{\psi_{A_{n+1}}}=\mcg_y\dket{\psi_{B_n}}=T_x\dket{\psi_{A_n}}.
\eea
In order for $\dket{\psi_{f.p.}}$ to describe a short-range-entangled (SRE) GSPT phase, each layer $\dket{\psi_{A_n}}$ in the fixed-point wavefunction must correspond to a SRE\footnote{We follow Kitaev's definition for short-range-entangled phases: a gapped phase is short-range entangled if it has no ground state degeneracy on any closed manifold.} 2d phase that preserves global symmetry $G_0$, with no ground state degeneracy on any closed manifold.

Therefore we have shown that any GSPT wavefunction, through a fully-symmetric finite depth quantum circuit (\ref{sym f.d.q.c.}), can be reduced to a fixed-point wavefunction (\ref{fixed point wf}). This fixed-point wavefunction describes an array of decoupled 2d layers ($A_n$ and $B_n$ in FIG. \ref{fig:glide1}) arranged in a glide-symmetric way. Although here we focus on 3d SRE phases with non-symmorphic glide symmetry, the above arguments and coupled layer construction can be easily generalized and applied to other spatial dimensions, and to non-symmorphic screw symmetries.

Although here we focus on SPT phases protected by glide symmetry, it's straightforward to see that all above discussions and fixed-point wavefunctions (\ref{fixed point wf}) equally apply to any 2-fold non-symmorphic symmetries, such as 2-fold screw symmetry where each layer is perpendicular to the screw axis. Another similar symmetry is the combination $\tilde T_x\equiv T_x\cdot\bst$ of Bravais lattice translation $T_x$ and time reversal $\bst$, which is preserved in anti-ferromagnetic topological insulators\cite{Mong2010,Mross2015} and superconductors\cite{Sahoo2016}.

\subsection{Classification of GSPT phases and anomalous surface topological orders}\label{sec:coupled layer}

As argued above, we have justified the coupled layer construction for any GSPT phase of interacting bosons/fermions. As illustrated in FIG. \ref{fig:glide1}, each layer $A_n/B_n$ is a SRE 2d symmetric phase $\dket{\psi_{A_n/B_n}}$. Due to glide symmetry condition (\ref{coupled layer:glide sym}), the SRE 2d layers $\{A_n,B_n\}$ are not independent of each other: once we fixed the SRE 2d phase $\dket{\psi_{A_0}}$ in one layer, all other layers and hence the 3d GSPT phase $\dket{\psi_{f.p.}}$ are automatically determined by glide symmetry. Therefore the classification of 3d GSPT phase $\dket{\psi}$ with symmetry group $G_s=\mbz^{\mcg_y}\times G_0$ is reduced to (but not equivalent to, as will be clear soon) the classification of 2d SRE phase $\dket{\psi_{A_0}}$ preserving on-site symmetry $G_0$. We call these 2d SRE phases $\dket{\psi_{A_0}}$ on layer $A_0$ as the ``\emph{2d root states}'' of 3d GSPT phases. We can therefore label a 3d GSPT phase $\dket{\Psi_0}_{\mcg_y}$ in (\ref{fixed point wf}) by its associated 2d SRE root phase $\dket{\psi_{A_0}}\simeq\dket{\Psi_0}$ by defining
\bea\label{fixed point wf:2d label}
\dket{\Psi_0}_{\mcg_y}\simeq\prod_n\dket{0}_{C_n\cup D_n}\otimes\dket{\psi_{A_n}=\Psi_0}\otimes\dket{\psi_{B_n}=\bar\Psi_0}
\eea
Here $\dket{\bar{\Psi}_0}$ is defined as the mirror-reflection image (equivalent to time reversal image) of 2d SRE phase $\dket{\Psi_0}$, so that 3d phase $\dket{\Psi_0}_{\mcg_y}$ defined above preserves glide symmetry $\mcg_y$.

Recent progress on SPT phases\cite{Senthil2015} leads to a classification of SRE phases with on-site symmetry group $G_0$. In particular, a SRE phase is gapped with no ground state degeneracy on any closed manifold. In 2d it can either a SPT phase\cite{Chen2013} with non-chiral edge modes; or an ``invertible'' phase\cite{Kong2014,Freed2014} with chiral edge modes, generated by the $E_8$ state of bosons\cite{Kitaev2006} and the $p_x+\imth p_y$ chiral superconductor of fermions\cite{Read2000}. The classification leads to an Abelian group structure of 2d SRE phases, where the addition of two elements in this Abelian group corresponds to the tensor product of two SRE phases (hence the Abelian addition rules).

Is there a one-to-one correspondence between 3d GSPT phases $\dket{\Psi_0}_{\mcg_y}$ in (\ref{fixed point wf:2d label}) and 2d SRE states $\dket{\Psi_0}$ with the same on-site symmetry? This naive expectation turns out to be wrong: in general \emph{the 3d GSPT classification is a subgroup of the 2d SRE classification with the same on-site symmetry group $G_0$}, for the following reason. While by definition two distinct 3d GSPT phases cannot share the same 2d SRE root state $\dket{\psi_{A_0}}\simeq\dket{\Psi_0}$ in their fixed-point wavefunctions, the reverse statement is not true. In other words, two different 2d SRE root phases $\dket{\Psi_0}$ and $\dket{\Psi_0^\prime}$ can lead to the same 3d GSPT phase $\dket{\Psi_0}_{\mcg_y}\simeq\dket{\Psi_0^\prime}_{\mcg_y}$. For a simplest example, consider a GSPT phase $\dket{\Psi_0}_{\mcg_y}$ with
\bea
\dket{\psi_{A_n}}\simeq\dket{\Psi_0},~~~\dket{\psi_{B_n}}\simeq\dket{\bar{\Psi}_0}
\eea
and another state $\dket{\bar\Psi_0}_{\mcg_y}$ with
\bea
\dket{\psi_{A_n}^\prime}\simeq\dket{\bar{\Psi}_0},~~~\dket{\psi_{B_n}^\prime}\simeq\dket{{\Psi}_0}
\eea
Clearly they describe the same GSPT phase, since they merely differ by a (arbitrary) choice of $\{A_n\}$ vs. $\{B_n\}$ layers. Therefore we have shown that two SRE 2d root phases that are mirror-reflection image of each other will lead to the same GSPT phase. Notice that 2d SRE phases (hence their 3d GSPT counterparts) generally satisfy the following addition rule:
\bea\label{2d SRE addition rule}
\dket{\Psi_0}\simeq \dket{\bar\Psi_0}\oplus\Big(\dket{\Psi_0}\oplus\dket{\Psi_0}\Big)\Longleftrightarrow
\dket{\Psi_0}\oplus\dket{\bar\Psi_0}\simeq\dket{0}.~~
\eea
where $\dket{0}$ represents the trivial product state. Now that we have argued $\dket{\Psi_0}_{\mcg_y}\simeq\dket{\bar\Psi_0}_{\mcg_y}$, an immediate consequence is the following ``$Z_2$ addition rule'' for any 3d GSPT phases:
\bea
&\notag \mathbf{Z_2}~\text{\bf addition rule of 3d GSPT phases:}\\
&\dket{\Psi_0}_{\mcg_y}\oplus\dket{\Psi_0}_{\mcg_y}\simeq\dket{0}_{\mcg_y}\label{z2 addition rule}
\eea
In other words, two copies of the same 3d GSPT phases always lead to the trivial product state. This important conclusion for GSPT phases allows us to further reduce 2d SRE classification and to achieve the 3d GSPT classification.

The above $Z_2$ addition rule (\ref{z2 addition rule}) for 3d GSPT phases also raises a question: given a fixed-point wavefunction (\ref{fixed point wf}) built from 2d SRE root phase $\dket{\psi_{A_0}}=\dket{\Psi_0}$, how do we know that it is a nontrivial GSPT phase (\ie it's not equivalent to the trivial product state $\dket{0}$) or not? To address this issue we explicitly construct certain topological orders, that symmetrically gap out the glide-invariant [001] side surface ($\hat x-\hat y$ plane in FIG. \ref{fig:glide1}) states of the GSPT phase. We further show that these surface topological orders (STOs) exhibit anomalous symmetry implementations\cite{Chen2016}, a fingerprint for interacting SPT phases\cite{Senthil2015} from bulk-boundary correspondence.

The coupled layer construction is particularly powerful for this purpose, when we study interaction effects on the [001] side surface which generally hosts glide-protected gapless surface states for weakly-interacting fermions (\eg ``hourglass fermions'' for $G_0=U(1)\times Z_4^\bst$). Specifically since each 2d layer intersects with the [001] side surface on its gapless edge, short-range interactions can couple these gapless 1d edge states to form a gapped surface topological order\cite{Kane2002,Teo2014,Mross2015,Mross2016,Sahoo2016,Song2016}. Such a ``coupled wire construction'' enables us to construct these 2d STOs and analyze their symmetry properties. The anomalies of STOs can be detected in physical responses of the gapped symmetric surface, such as thermal and electric Hall conductance, or the magnetic flux (in particular $\pi$ flux) on the surface. In contrary, the side surface states of a trivial GSPT phase can always be gapped out symmetrically without developing any topological orders. This will be demonstrated in all examples.

\begin{figure}[ht]
\begin{center}
\includegraphics[width=8cm]{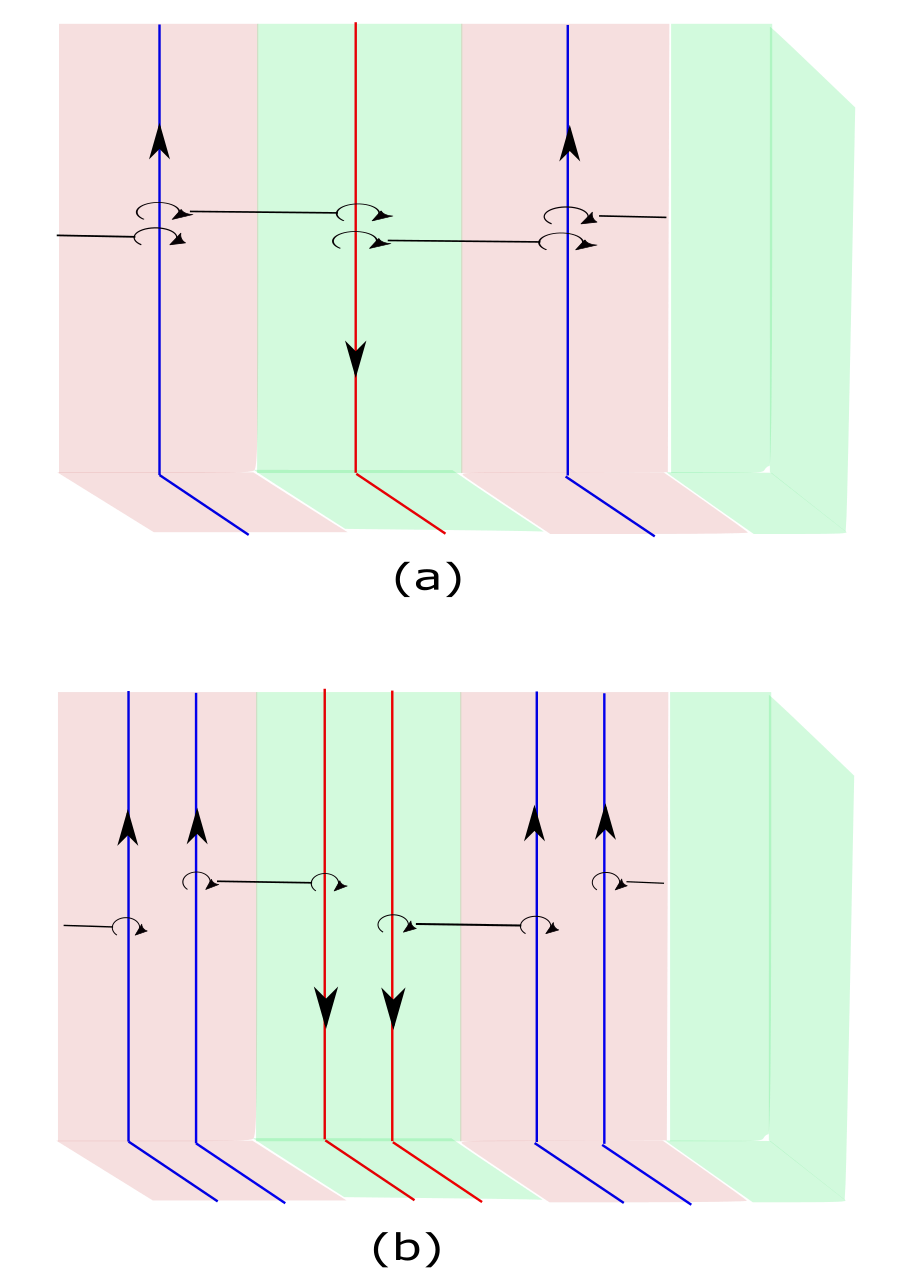}
\end{center}
\caption{(a) Surface states on the glide-invariant [001] side surface of a 3d GSPT phase $\gspt{\Psi_0}$ cannot be trivially gaped out without breaking any symmetry. Each red/blue line corresponds to the edge states of each layer at the intersection with the side surface. Proper couplings between two neighboring edges can lead to anomalous STOs that preserves all symmetries, such as the case of $\gspt{E_8}$ state in section \ref{sec:BG}. (b) The surface states of the trivial state $\dket{\Psi_0}_{\mcg_y}\oplus\dket{\Psi_0}_{\mcg_y}$ can be trivially gapped out in a pairwise fashion without breaking any symmetries, demonstrating the $Z_2$ addition rule (\ref{z2 addition rule}) of 3d GSPT phases.}
\label{fig:glide2}
\end{figure}

\subsection{Coupled wire construction of surface topological orders}\label{sec:coupled wire}

As mentioned earlier, we will use the ``coupled wire construction''\cite{Kane2002,Teo2014} to study anomalous STO on the [001] side surface of a GSPT phase, where each wire is simply the edge states at the intersection of [001] side surface and
each 2d SRE layer in the coupled layer construction of 3d GSPT phases. These gapless 1d edge states of 2d SRE bosonic/fermionic phases have been studied extensively in the literature\cite{Lu2012a,Levin2012a}. To construct the anomalous STOs, our strategy is to either couple these ``edge wires'' directly, or to deposit extra 1d quantum wires on the side surface and couple them together with the ``edge wires''. This will be demonstrated in the examples.

This coupled wire construction for GSPT surface states also provides an alternative argument for the $Z_2$ addition rule (\ref{z2 addition rule}) for any bosonic/fermionic GSPT phase, as illustrated in FIG. \ref{fig:glide2}(b). Consider a 3d GSPT phase $\dket{\Psi_0}_{\mcg_y}\oplus\dket{\Psi_0}_{\mcg_y}$, where the protected edge modes of each 2d SRE phase $\dket{\Psi_0}$ are labeled by one arrowed blue line in FIG. \ref{fig:glide2}(b). The edge modes of its mirror-reflection (or time-reversal) image $\dket{\bar\Psi_0}$ are labeled by one red line with an opposite arrow in FIG. \ref{fig:glide2}(b). Due to addition rule (\ref{2d SRE addition rule}) for a pair of 2d SRE phases $\{\dket{\Psi_0},\dket{\bar\Psi_0}\}$ that are time-reversal (or mirror-reflection) counterparts of each other, their edge states together (one red plus one blue lines) can be gapped out symmetrically. More concretely, the robust edge states of a 2d SRE phase $\dket{\Psi_0}$ can be described by chiral boson fields $\{\phi^I\}$
\bea\label{edge:spt}
\mathcal{L}_{edge}={4\pi}\sum_{I,J}\partial_t\phi^I(y,t){\bf K}_{I,J}\partial_y\phi^J(y,t)+\cdots
\eea
where $\cdots$ represents non-universal kinetic energy terms, with commutation relation
\bea\label{commutator:spt edge}
[\phi^I(y^\prime),\partial_y\phi^J(y)]=2\pi\imth{\bf K}^{-1}_{I,J}\delta(y-y^\prime).
\eea
Therefore the gapless surface states on glide-invariant [001] side surface of state $\dket{\Psi_0}_{\mcg_y}\oplus\dket{\Psi_0}_{\mcg_y}$ are described by
\bea
\mathcal{L}_0=\sum_{n\in\frac\mbz2}\sum_{a=1,2}\frac{(-1)^{2n}}{4\pi}\sum_{I,J}\partial_t\phi_{n,a}^I(y,t){\bf K}_{I,J}\partial_y\phi_{n,a}^J(y,t)+\cdots\notag
\eea
In particular under glide symmetry operation, the chiral bosons transform as
\bea
\phi^I_{n,a}(y)\overset{\mcg_y}\longrightarrow\phi^I_{n+1/2,a}(-y)
\eea
Since $\{\phi^I_n\}$ transform in the same way under global symmetries for $\forall~n,a$, the following coupling terms that pair up two neighboring edges
\bea\label{interwire:z2 addition rule}
\mathcal{H}_{int}=-C_0\sum_{n\in\mbz/2}\sum_I\cos(\phi^I_{n,2}-\phi^I_{n+\frac12,1})
\eea
trivially gap out the surface states without breaking any symmetry. Therefore the glide-invariant [001] surface states of 3d GSPT phase $\dket{\Psi_0}_{\mcg_y}\oplus\dket{\Psi_0}_{\mcg_y}$ can be gapped out without breaking any symmetry, in a pairwise fashion illustrated in FIG. \ref{fig:glide2}(b). Due to the bulk-boundary correspondence, this 3d GSPT state is equivalent to the trivial product state, reaffirming the $Z_2$ addition rule (\ref{z2 addition rule}).

Previously in the coupled layer construction of certain ``weak'' SPT phases, their symmetric STOs have been constructed by adding ``plates'' of 2d topological orders on the side surface between two neighboring layers, then gapping out the edge states of plates and the layers by certain inter-wire couplings\cite{Mross2015,Mross2016,Song2016}. However in this approach the side surface is by construction not a uniform 2d system, where it's subtle to even define a topological order. This concern motivates us to follow a different approach: instead of depositing 2d plates on the side surfaces, we deposit 1d quantum wires on the side surface between two neighboring layers in a glide-invariant fashion. We further construct symmetric 2d STOs by properly coupling the 1d wires with gapless edge states of each layer.

Below we briefly review the coupled wire construction of 2d topological orders, focusing on Abelian $Z_k$ gauge theories that will be frequently encountered in STOs of various GSPT phases. We first show how to construct a 2d $Z_k$ gauge theory by properly coupling a 2d array of boson quantum wires, and then demonstrate the bulk and edge anyons and their fractional statistics in the $Z_k$ gauge theory. The basic ideas and strategy are explained in Appendix \ref{sec:anyon}. Consider a 2d array of 1d Luttinger liquids $\{\varphi_l,\theta_l|l\in\mbz\}$ described by
\bea
&\notag\mathcal{L}_{Z_k}=\mathcal{L}_0+\mathcal{H}_{int},\\
&\label{lag:zk gauge theory}\mathcal{L}_0=\frac1{2\pi}\sum_{l\in\mbz}\partial_t\varphi_l\partial_y\theta_l+\cdots
\eea
with commutation relation
\bea
&\notag[\theta_{l_1}(y_1),\varphi_{l_2}(y_2)]=[\varphi_{l_1}(y_1),\theta_{l_2}(y_2)]\\
&=\imth\pi\delta_{l_1,l_2}\text{Sign}(y_1-y_2)\label{commutator:spin chain}
\eea
Physically each quantum wire can be realized by \eg the Luttinger liquid of a 1d gapless ``XYX'' spin chain $\hat H_{XYX}=-J_1\sum_r(S^x_rS^x_{r+1}+S^z_rS^z_{r+1})$, where physical observables can be expressed in terms of boson fields $\{\varphi(r),\theta(r)\}$ as
\bea\label{chiral boson:spin chain}
S^+\equiv S^z+\imth S^x\sim e^{\imth\varphi},~~~\rho_{S^y}\sim\frac{\partial_r\theta(r)}{2\pi}.
\eea
Clearly under on-site $U(1)$ charge rotation ($\hat Q$ labels the total charge) and time reversal operation, the boson fields $\{\varphi_l,\theta_l|l\in\mbz\}$ transform as
\bea\label{sym:U(1)+trs:spin chain}
\bpm\varphi_l\\ \theta_l\epm\overset{e^{\imth\alpha\hat Q}}\longrightarrow\bpm\varphi_l\\ \theta_l\epm,~~~\bpm\varphi_l\\ \theta_l\epm\overset{\bst}\longrightarrow\bpm\varphi_l+\pi\\-\theta_l\epm
\eea
The following inter-wire terms
\bea\label{zk gauge theory:interwire}
\mathcal{H}_{int}=-\sum_{l\in\mbz}C_l\cos\hat\Theta_l,~~~\Theta_l=\varphi_{l-1}-k\theta_{l}-\varphi_{l+1}.
\eea
gap out all wires and lead to an Abelian topological order, described by Abelian Chern-Simon theory with ${\bf K}$ matrix
\bea
{\bf K}_{Z_k}=\bpm0&k\\k&0\epm
\eea
This is the coupled wire construction of 2d $Z_k$ gauge theory.

The anyons in this 2d topological order can be identified both on the edge or in the bulk. To see this, we first consider an open boundary between the $(2l-1)$-th and $(2l)$-th wire by setting $C_{2l-1}=C_{2l}=0$. Zero-modes on the right (or left) edge, \ie vertex operators commuting with all cosine terms correspond to the anyons on the edge, in this case
\bea
&\notag e^{\imth\phi_l^e}=e^{\imth\varphi_{2l}/k}\sim e^{\imth(\frac1k\varphi_{2l-2}-\theta_{2l-1})},\\
&e^{\imth\phi_l^m}=e^{\imth(\theta_{2l}+\frac1k\varphi_{2l+1})}\sim e^{\imth\varphi_{2l-1}/k}.\label{anyon:zk gauge theory}
\eea
We use $e$ and $m$ to label the gauge charge and gauge flux of a $Z_k$ gauge theory, and their fractional (mutual) statistics is indicated by commutation relation
\bea\label{commutator:zk gauge theory}
[\phi_l^e(y_1),\phi_l^m(y_2)]=\imth\frac{\pi}{k}\text{Sign}(y_1-y_2).
\eea
Meanwhile on a closed manifold with no boundary ($C_l\neq0,~\forall~l$), anyons $e^{\imth\phi^{e/m}_l}$ in (\ref{anyon:zk gauge theory}) are nothing but kinks\cite{Teo2014} of cosine terms $C_{2l}$ and $C_{2l-1}$, which create $2\pi$ phase slips in the arguments $\Theta_{2l},\Theta_{2l-1}$ of associated cosine terms. One anyon of type $a$ can hop from one wire to another across the bulk, via the following hopping operators
\bea
T^a_{l_1,l_2}=\phi_{l_2}^a-\phi_{l_1}^a+\sum_lt_l\Theta_l,~~~t^a_l\in\mathbb{R}.
\eea
consisting of a string of local operators. For example in our case of $Z_k$ gauge theory (\ref{zk gauge theory:interwire}), the anyon hopping operators are
\bea
&\notag T^e_{l,l+1}=-\theta_{2l+1},~~~t_{2l+1}^e=\frac1k;\\
&T^m_{l,l+1}=-\theta_{2l},~~~t_{2l+2}^m=\frac1k.
\eea
As discussed in Appendix \ref{sec:anyon}, the braiding statistics of anyons can be easily computed based on the anyon hopping operators, and the results are consistent with (\ref{commutator:zk gauge theory}) for the edge anyons.

The construction of STO is very similar to (\ref{zk gauge theory:interwire}) for 2d $Z_k$ gauge theory, where anyons and their statistics can be identified as reviewed in Appendix \ref{sec:anyon} and demonstrated above. What makes the STO anomalous is one important difference: a part (sometimes all) of the 1d quantum wires $\{\varphi_l,\theta_l\}$ in (\ref{lag:zk gauge theory}) will be replaced by the protected edge states of nontrivial 2d SRE phases, whose symmetry implementations cannot be realized in a pure 1d system. This replacement is crucial for realizing an anomalous STO respecting glide symmetry, and will be encountered in many examples.\\

%
%

While the strategy established in this section applies to all global symmetries, in this paper we mostly focus on examples with on-site symmetry group $G_0$ generated by $U(1)$ charge/spin symmetry and/or time reversal symmetry  $\bst$. In particular, we consider 5 on-site symmetry groups for bosonic GSPT phases in section \ref{sec:boson}
\bea\label{sym:no}
\text{No on-site symmetry:}~~G_0=Z_1
\eea
\bea\label{sym:trs}
\text{Time reversal symmetry:}~~G_0=Z_2^\bst
\eea
\bea\label{sym:U(1)}
U(1)~\text{symmetry:}~~G_0=U(1)
\eea
\bea\label{sym:trs+U(1)}
U(1)~\text{charge and time reversal:}~~G_0=U(1)\rtimes Z_2^\bst
\eea
\bea\label{sym:trs+U(1) spin}
U(1)~\text{spin and time reversal:}~~G_0=U(1)\times Z_2^\bst
\eea
for 3d GSPT phases of interacting bosons. The results are summarized in TABLE \ref{tab:a}. For all bosonic GSPT phases studied here, it turns out that their surface states can always be gapped symmetrically by \emph{Abelian} $Z_2$ topological orders, with 4-fold ground state degeneracy on torus. The anomalies of these STOs are also summarized in TABLE \ref{tab:a}.

The situation is more complicated for interacting fermions, since the microscopic fermions can either be half-integer-spin Kramers doublets with $\bst^2={\bold P}_f$ where ${\bold P}_f=(-1)^{\hat F}$ represents the fermion parity; or integer-spin Kramers singlets with $\bst^2=1$. Therefore the above 5 cases for interacting bosons will lead to 7 cases for interacting fermions discussed in section \ref{sec:F}, where the on-site symmetry can be categorized into Altland-Zirnbauer (AZ) 10-fold way\cite{Altland1997}. The corresponding results are summarized in TABLE \ref{tab:b}. Unlike in bosonic GSPT case, in certain fermionic GSPT phases, \emph{non-Abelian} topological orders with a non-integer chiral central charge $c_-$ of their edge states are necessary to symmetrically gap out the surface states. Examples with these non-Abelian STOs include topological superconductors in AZ symmetry class $D$ and $DIII$, and topological insulator in class $A$ (see TABLE \ref{tab:b}).

\section{Bosonic GSPT phases}\label{sec:boson}

We first discuss GSPT phases of interacting bosons, following the strategy described in the previous section. The 5 cases (\ref{sym:no})-(\ref{sym:trs+U(1) spin}) will be discussed separately in 5 subsections. All results are briefly summarized in TABLE \ref{tab:a}.


%
%
%

\begin{table*}
\centering
\begin{tabular}{ | c || c | c || c | c  |}
    \hline
     \multirow{2}{3cm}{On-site symmetry group $G_0$} & \multirow{2}{2.5cm}{Classification of 2d SRE phases} & 2d SRE root states & \multirow{2}{3cm}{Classification of 3d GSPT phase}&\multirow{2}{3.5cm}{Anomaly of STOs ($Z_2$ topological orders)}\\
     &&&&\\ \hline
    $Z_1$ & $\mathbb{Z}$ & $E_{8}$ & $\mathbb{Z}_{2}$&e$f$m$f$ \\ \hline
    $U(1)$ & $\mathbb{Z}\times\mathbb{Z}$ & BIQH/neutral~$E_{8}$ &$\mathbb{Z}_{2}\times\mathbb{Z}_{2}$&e$C$m$C$/e$f$m$f$ \\ \hline
    $Z_2^\bst$ & $\mathbb{Z}_{1}$ & None  & $\mathbb{Z}_{1}$&None \\ \hline
    $U(1)\rtimes Z_2^\bst$ & $\mathbb{Z}_{2}$& BQSH & $\mathbb{Z}_{2}$& e$\mcg_y\bst$m$C$(glide action on $\pi$ flux) \\
    \hline
        $U(1)\times Z_2^\bst$ & $\mathbb{Z}_{1}$& None & $\mathbb{Z}_{1}$& None \\
    \hline
            $Z_N$ & $\mathbb{Z}_{N}\times\mbz$& $Z_N$ SPT/$E_8$& $\mbz_{(N,2)}\times\mbz_2$ & e$C$m$C$/e$f$m$f$ \\
    \hline
        $Z_N\rtimes Z_2^\bst$ & $\mathbb{Z}_{(N,2)}^2$& $Z_N\rtimes Z_2^T$ SPT & $\mathbb{Z}_{(N,2)}^2$& e$\mcg_y\bst$m$C$/e$\mcg_yC$m$C$\\
        \hline
                $Z_N\times Z_2^\bst$ & $\mathbb{Z}_{(N,2)}^2$& $Z_N\times Z_2^T$ SPT & $\mathbb{Z}_{(N,2)}^2$& e$\mcg_y\bst$m$C$/e$\mcg_yC$m$C$\\
    \hline
\end{tabular}
\caption{Summary of 3d boson GSPT phases with various on-site symmetry group $G_0$. The fixed-point wavefunction (\ref{fixed point wf:2d label}) of a GSPT phase can be constructed by coupling layers of 2d SRE phases with the same on-site symmetry, allowing us to classify 3d GSPT phases and study their anomalous surface topological orders (STOs). We use $(a,b)$ to denote the greatest common divisor of two integers $a$ and $b$.}
\label{tab:a}
\end{table*}

\subsection{No on-site symmetry: $\mbz_2$ classification}\label{sec:BG}

In the absence of any on-site symmetry, 2d SRE phases of bosons are classified and characterized by an integer-valued index $\nu\in\mbz$. The ``generator'' of this integer group, \ie corresponding 2d root state is the $E_8$ state\cite{Kitaev2006,Lu2012a}. The definitive physical properties of $E_8$ state is quantized thermal Hall conductance\cite{Kane1997} $\kappa_H/T=\frac{\pi^2}{3}\frac{k_B^2}{h}c_-$ with chiral central charge $c_-=8$. In particular the chiral edge states of $E_8$ state is described by Lagrangian density
\bea\label{edge:E8}
&\mathcal{L}_{edge}=\frac1{4\pi}\sum_{I,J}\partial_t\phi^I(y,t){\bf K}^{I,J}_{E_8}\partial_y\phi^J(y,t)+\cdots,\\
&{\bf K}_{E_8}=\bpm2&-1&0&0&0&0&0&0\\-1&2&-1&0&0&0&0&0\\0&-1&2&-1&0&0&0&0\\0&0&-1&2&-1&0&0&0\\0&0&0&-1&2&-1&0&-1\\0&0&0&0&-1&2&-1&0\\
0&0&0&0&0&-1&2&0\\0&0&0&0&-1&0&0&2\epm=\Gamma_8\cdot\Gamma_8^T\notag
\eea
where ``$\cdots$'' represents non-universal kinetic energy terms. ${\bf K}_{E_8}$ is the Cartan matrix of Lie group $E_8$ and
\bea
\Gamma_8=\bpm1&-1&0&0&0&0&0&0\\0&1&-1&0&0&0&0&0\\0&0&1&-1&0&0&0&0\\0&0&0&1&-1&0&0&0\\0&0&0&0&1&-1&0&0\\0&0&0&0&0&1&1&0\\
-\frac12&-\frac12&-\frac12&-\frac12&-\frac12&-\frac12&-\frac12&-\frac12\\0&0&0&0&0&1&-1&0 \epm
\eea
is the simple root matrix of $E_8$ group. $\{\phi^I\}$ are chiral boson operators satisfying commutation relation
\bea
[\phi^I(y^\prime),\partial_y\phi^J(y)]=2\pi\imth{\bf K}^{-1}_{I,J}\delta(y-y^\prime).
\eea

According to $Z_2$ addition rule (\ref{z2 addition rule}), we know that two copies of $E_8$ states ($\nu=\pm2$) in each layer lead to the trivial product state \ie $\dket{E_8}_{\mcg_y}\oplus\dket{E_8}_{\mcg_y}\simeq\dket{0}_{\mcg_y}$. It turns out a single $E_8$ state in each layer leads to a nontrivial GSPT phase $\dket{E_8}_{\mcg_y}\neq\dket{0}_{\mcg_y}$, resulting in a $\mbz_2$ classification of GSPT phases with no on-site symmetry.

Below we explicitly construct anomalous STOs on the [001] side surface of $\dket{E_8}_{\mcg_y}$ state, in the framework of coupled wire construction. We label the $E_8$ chiral bosons from the ``edge wire'' of $A_n$ layer by $\{\phi^I_n|n\in\mbz\}$, and those of $B_n$ layer by $\{\phi^I_n|n\in\mbz+\frac12\}$. The Lagrangian density for decoupled wires on [001] side surface is given by
\bea
\mathcal{L}_0=\sum_{n\in\mbz/2}\frac{(-1)^{2n}}{4\pi}\sum_{I,J=1}^8\partial_t\phi_n^I(y,t){\bf K}^{I,J}_{E_8}\partial_y\phi_n^J(y,t)+\cdots
\eea
Clearly under glide symmetry operation, the chiral bosons transform as
\bea\label{glide sym:E8}
\phi^I_n(y)\overset{\mcg_y}\longrightarrow\phi^I_{n+1/2}(-y)
\eea
The root matrix $\Gamma_8$ defines a basis change of chiral bosons\cite{Cano2014,Mross2015} with simple commutation relations:
\bea
&\tilde\phi^I_n\equiv\sum_J(\Gamma_8)_{J,I}\phi^J_n,\\
\notag&[\tilde\phi^I_n(y^\prime),\partial_y\tilde\phi^J_m(y)]=(-1)^{2n}2\pi\imth\delta_{I,J}\delta_{m,n}\delta(y-y^\prime)
\eea
In other words, each column of matrix $\Gamma_8$ corresponds to a chiral fermion mode on the edge of a $\sigma_{xy}=1$ integer quantum Hall (IQH) state. These ``chiral fermions'' are however non-local operators due to the non-integer ($-\frac12$) entries in matrix $\Gamma_8$. To construct a fully-symmetric STO out of these ``edge wires'', our strategy is to couple the chiral fermions $\{\tilde\phi^I_n\}$ in pairs between two neighboring wires, as illustrated in FIG. \ref{fig:glide2}(a). Specifically we consider the following glide-invariant inter-wire couplings (with positive strength $C_I>0$):
\bea\label{interwire term:E8:efmf}
\mathcal{H}_{int}=-\sum_{n\in\mbz/2}\sum_{I=1}^4C_I\cos\Big[\sum_{J=1}^4{\bf M}_{I,J}(\tilde\phi^{J}_n+\tilde\phi^{J+4}_{n+\frac12})\Big]
\eea
which pair up counter-propagating chiral fermions $\tilde\phi^J_n$ and $\tilde\phi^{J+4}_{n+\frac12}$. A proper choice of $4\times4$ matrix ${\bf M}$
\bea
{\bf M}=\bpm1&1&0&0\\1&-1&0&0\\0&1&-1&0\\0&0&1&-1\epm
\eea
can make (\ref{interwire term:E8:efmf}) a local Hamiltonian that fully gap out the surface while preserving glide symmetry (\ref{glide sym:E8}). Note that all cosine terms commute with each other, and can be minimized simultaneously. The argument of each cosine term is hence pinned at certain classical minimum in the the ground state. Bulk anyons correspond to kinks of these cosine terms\cite{Teo2014}. Following the strategy outlined in Appendix \ref{sec:anyon}, it's straightforward to identify the inequivalent bulk anyons:
\bea\notag
e\sim e^{\frac\imth2\sum_{I=1}^4\tilde\phi^I_n}=e^{\imth(\frac{\phi^4_n}2-\phi^7_n)},\\
\notag m\sim e^{\frac\imth2(\sum_{I=2}^4\tilde\phi^I_n-\tilde\phi^1_n)}=e^{\imth(\frac{\phi^4_n-\phi^7_n}2-\phi^1_n)},\\
\epsilon\sim e^{\imth\tilde\phi^4_I}=e^{\imth(\phi^4_n-\phi^3_n-\frac{\phi^7_n}2)}.
\eea
These Abelian anyons can also be obtained from the gapless edge states of the STO, by cutting an edge between $A_n$ and $B_n$ layer. All 3 anyons obey fermi self-statistics and satisfy the following $Z_2\times Z_2$ fusion rule:
\bea
&e\times e=m\times m=1,\\
&\epsilon=e\times m.\label{fusion rule:z2}
\eea
They also obey mutual semion statistics, \ie $e^{\imth\pi}=-1$ phase when braiding $e$ around $m$ once. Such a STO is known as $Z_2$ topological order ``e$f$m$f$''\cite{Vishwanath2013,Wang2013a}, described by the following ${\bf K}$ matrix (Cartan matrix of Lie group $SO(8)$)
\bea
{\bf K}_{SO(8)}=\bpm2&-1&-1&-1\\-1&2&0&0\\-1&0&2&0\\-1&0&0&2\epm
\eea
in the Abelian Chern-Simons theory framework. A pure 2d e$f$m$f$ phase have chiral central charge $c_-=4$ and necessarily breaks orientation-reversing glide (and PT) symmetry. Therefore our glide-invariant e$f$m$f$ STO is anomalous and can only be realized on the surface of 3d GSPT phase $\dket{E_8}_{\mcg_y}$. Therefore we have justified that $\gspt{E_8}$ is a nontrivial GSPT phase, and hence the $\mbz_2$ classification of GSPT phase with no on-site symmetry.

\subsection{Time reversal symmetry: $\mathbb{Z}_{1}$ classification}

With on-site time reversal symmetry, $E_8$ states are forbidden and 2d bosonic SRE phases have a trivial classification\cite{Chen2013} $H^3(Z_2^\bst,U(1))=\mbz_1$. With no nontrivial 2d root phases, the classification of 3d GSPT phases with $G_0=Z_2^\bst$ is hence also trivial, as shown in TABLE \ref{tab:a}.

%
%

\begin{figure}[ht]
\begin{center}
\includegraphics[width=8cm]{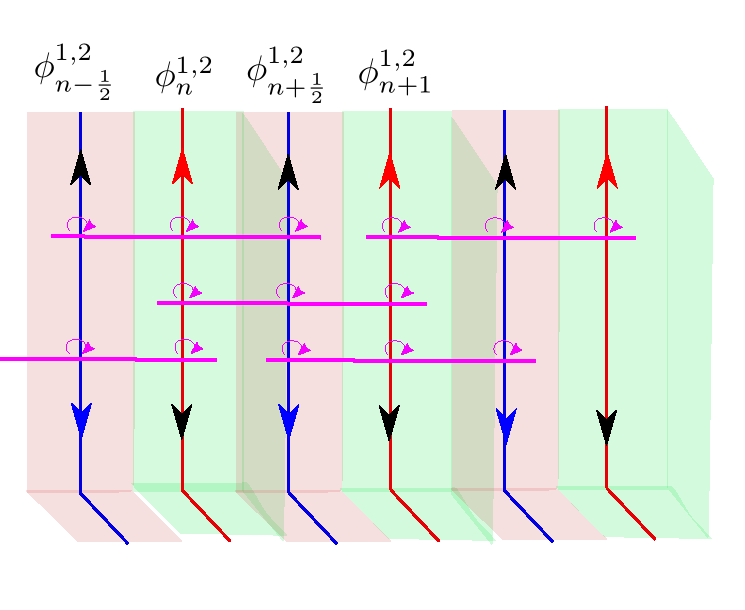}
\end{center}
\caption{Illustration of interwire couplings (\ref{biqh:interwire}) that realize the symmetric STO e$C$m$C$ on the [001] side surface of 3d GSPT phase $\gspt{\text{BIQH}}$, see section \ref{sec:BGU}.}
\label{fig:glide3}
\end{figure}

\subsection{$U(1)$ symmetry: $\mathbb{Z}_{2}\times\mathbb{Z}_{2}$ classification} \label{sec:BGU}


In the presence of on-site $U(1)$ charge (or spin) symmetry, 2d SRE boson phases have a $\mbz\times\mbz$ classification\cite{Lu2012a}, labeled by a pair of integers $\{\nu\in\mbz,q\in\mbz\}$. Physically the pair of integers correspond to chiral central charge and Hall conductance of a 2d SRE phase:
\bea
c_-=8\nu,~~~~~\sigma_{xy}=2q\frac{Q_0^2}{h}
\eea
where $Q_0$ is the fundamental charge carried by one boson. Physically a state $\{\nu,q\}$ can be obtained by stacking $\nu$ copies of $E_8$ states of charge-neutral bosonic excitons, and $q$ copies of bosonic integer quantum Hall (BIQH) states\cite{Lu2012a,Senthil2013} of charge-$Q_0$ bosons. These 2d SRE phases labeled as $[\nu,q]$ satisfy the following addition rule:
\bea
[\nu_1,q_1]\oplus[\nu_2,q_2]=[\nu_1+\nu_2,q_1+q_2].
\eea
Two generators of these 2d $U(1)$-symmetric SRE phases are the neutral $E_8$ state $[\nu=1,q=0]$ and the BIQH state $[\nu=0,q=1]$. Their associated 3d GSPT phases are hence generated by $\gspt{1,0}=\gspt{E_8}$ and $\gspt{0,1}=\gspt{\text{BIQH}}$, satisfying $Z_2\times Z_2$ fusion rule (\ref{z2 addition rule}) \ie
\bea
\gspt{\nu\in2\mbz,q}\simeq\gspt{\nu,q\in2\mbz}\simeq\gspt{0}.
\eea
Previously in section \ref{sec:BG} we have shown that $\gspt{E_8}=\gspt{1,0}$ is a nontrivial GSPT phase with anomalous STOs like e$f$m$f$. Below we show that $\gspt{\text{BIQH}}=\gspt{0,1}$ is also a nontrivial GSPT phase by constructing its anomalous STO, and therefore establish the $\mbz_2\times\mbz_2$ classification of 3d GSPT phases with on-site $U(1)$ symmetry.

The edge states of 2d BIQH root state $\dket{\text{BIQH}}=\dket{0,1}$ is described by\cite{Lu2012a} chiral bosons $\{\phi^I|I=1,2\}$
\bea\label{edge:biqh}
\mathcal{L}_{edge}=\frac1{4\pi}\sum_{I,J=1,2}\partial_t\phi^I{\bf K}_{I,J}\partial_y\phi^J+\cdots,
\eea
with ${\bf K}$ matrix and charge vector ${\bf q}$
\begin{equation}
\bold{K}=\left(\begin{array}{cc} 0 & 1\\
1 & 0 \end{array}\right),~~~\bold{q}_{\text{BIQH}}= \left(\begin{array}{c} 1\\
1 \end{array}\right)
\end{equation}
The chiral bosons transform under $U(1)$ charge/spin rotation (by angle $\alpha$) as
\bea\label{sym:U(1):biqh}
\vec\phi\equiv\bpm\phi^1\\ \phi^2\epm\overset{e^{\imth\alpha\hat Q}}\longrightarrow\vec\phi+\alpha\cdot{\bf K}^{-1}{\bf q}_{\text{BIQH}}=\bpm\phi^1+\alpha\\ \phi^2+\alpha\epm
\eea

\begin{figure}[ht]
\begin{center}
\includegraphics[width=8cm]{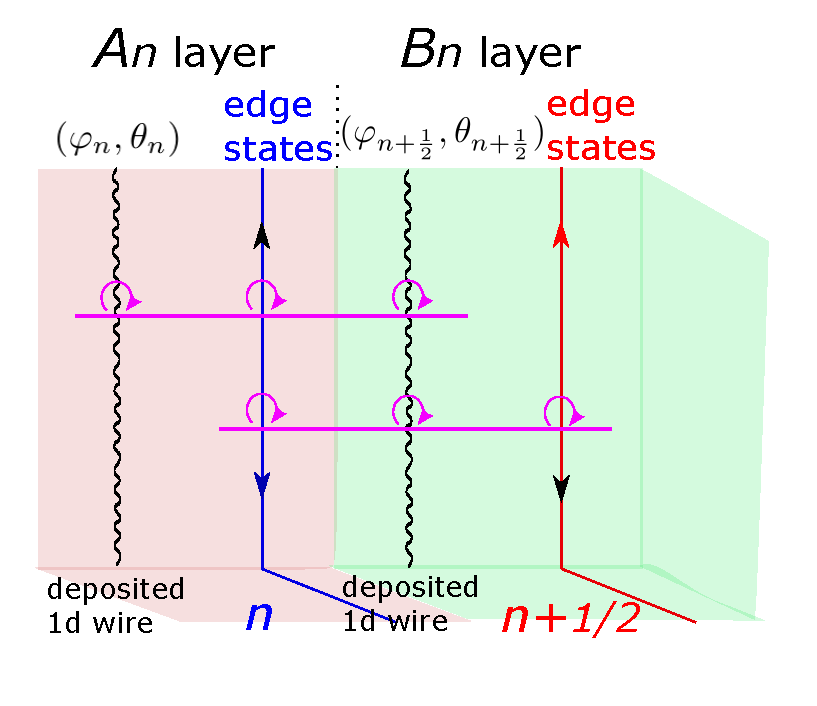}
\end{center}
\caption{Coupled wire construction of anomalous STO e$\mcg_y\bst$m$C$ for 3d GSPT phases with both $U(1)$ and time reversal symmetries. One spin chain (\ref{commutator:spin chain})-(\ref{chiral boson:spin chain}) depicted by a wiggled line is deposited between two neighboring 2d bosonic/fermonic QSH layers, whose edge states are labeled by the red and blue arrowed lines. The interwire terms couple 3 neighboring 1d Luttinger liquids together, stablizing the STO without breaking any symmetry.}
\label{fig:glide4}
\end{figure}

Now we consider the glide-invariant [001] side surface of 3d GSPT phase $\gspt{\text{BIQH}}=\gspt{\nu=0,q=1}$. Gapless edge states of BIQH states will appear at the intersection of each 2d plane $\{A_n,B_n\}$ in (\ref{fixed point wf:2d label}) and the side surface. We label these gapless chiral bosons from $A_n$ layer by $\{\phi^I_{n}|n\in\mbz\}$, and those from $B_n$ layer by $\{\phi^I_{n+\frac12}|n\in\mbz\}$, illustrated by the ``edge wires'' in FIG. \ref{fig:glide4}. These gapless surface states are described by the following glide-invariant Lagrangian density
\bea\label{biqh:edge wires}
\mathcal{L}_0=\sum_{n\in\mbz/2}\frac{(-1)^{2n}}{4\pi}\sum_{I,J}\partial_t\phi^I_n{\bf K}_{I,J}\partial_y\phi^J_n+\cdots
\eea
with commutation relation
\bea\label{commutator:biqh}
[\phi^I_m(x),\phi^J_n(y)]=(-1)^{2n}\pi\imth{\bf K}^{-1}_{I,J}\delta_{m,n}\text{Sign}(x-y).
\eea

The anomalous STO on side surface of $\gspt{\text{BIQH}}$ is realized by the following coupled wire construction:
\bea\label{biqh:sto}
\mathcal{L}_{STO}=\mathcal{L}_0+\mathcal{H}_{int}
\eea
where $\mathcal{L}_{0}$ is given in (\ref{biqh:edge wires}), and the inter-wire coupling terms are given by
\bea
\mathcal{H}_{int}=C_0\sum_{n\in\frac\mbz2}\cos(\hat L_n),~~~\hat L_n=\phi^1_{n-\frac12}-2\phi^2_n+\phi^1_{n+\frac12}.\label{biqh:interwire}
\eea
where $C_0$ is a real constant. They preserve both on-site $U(1)$ and glide symmetries. These inter-wire terms are illustrated in FIG. \ref{fig:glide3}, where every 3 neighboring wires are coupled together. Again all cosine terms commute with each other and hence can be minimized simultaneously.

What type of STO is developed out of inter-wire couplings (\ref{biqh:interwire})? This issue can be addressed in two approaches. One way is to identify the 2d ``bulk'' anyons in the above coupled-wire model and their hopping operators, as outlined in Appendix \ref{sec:anyon}. One can easily identify 3 types of inequivalent anyons:
\bea\label{anyon:eCmC}
&e\sim e^{\imth\phi^e_n}=e^{\imth\phi_n^1/2},~~~m\sim e^{\imth\phi^m_n}=e^{\imth(\frac12\phi_{n+\frac12}^1-\phi^2_n)},\\
\notag&\epsilon=e\times m\sim e^{\imth(\phi^e_n+\phi^m_n)}=e^{\imth(\frac{\phi_{n+\frac12}^1+\phi_n^1}2-\phi_n^2)},~~n\in\mbz.
\eea
that can freely hop in the STO. From the $U(1)$ symmetry transformation rules (\ref{sym:U(1):biqh}), both $e$ and $m$ anyons carry half-integer charges of $U(1)$ symmetry. It's also straightforward to show that they satisfy the $Z_2\times Z_2$ fusion rules in (\ref{fusion rule:z2}), and mutual semion statistics between two different anyons. Both $e$ and $m$ obey bose self-statistics, while $\epsilon$ obeys fermi self-statistics. Another approach to identify the STO is to consider a fictitious (since the surface of a 3d system is not ``edgable'') open edge of the 2d STO, by cutting a boundary between any two wires. For example for the boundary between wires $\phi_{n-\frac12}^{1,2}$ and $\phi_{n}^{1,2}$, the gapless edge modes on the right edge are given by $\{\phi_n^e,\phi_n^m\}$ in (\ref{anyon:eCmC}). The ${\bf K}$ matrix and charge vector ${\bf q}$ for the STO are given by
\bea
{\bf K}_{STO}=-\bpm0&2\\2&0\epm,~~~{\bf q}_{STO}=\bpm1\\-1\epm\mod2.
\eea
The charge vector is defined only$\mod2$, since one can always redefine the anyon $e$ (or $m$) by combining a local boson with it, which changes the anyon charge by 1 without affecting fractional statistics.

Therefore the STO on top of GSPT phase $\gspt{\text{QHE}}$ is a $Z_2$ topological order of toric code type\cite{Kitaev2003}. The anomaly of this toric code STO lies in its non-vanishing Hall conductance:
\bea
\sigma_{xy}^{STO}={\bf q}_{STO}^T{\bf K}^{-1}_{STO}{\bf q}_{STO}=1\mod2\neq0
\eea
Such a PT-breaking nonzero Hall conductance contradicts the orientation-reversing glide symmetry, and hence cannot be realized in a pure 2d system. It characterizes the anomalous glide symmetry implementation on the [001] side surface of GSPT phase $\gspt{\text{BIQH}}$. This anomalous STO is known as e$C$m$C$\cite{Vishwanath2013,Wang2013a} in the context of 3d SPT phase with $U(1)\rtimes Z_2^\bst$ symmetry.

As mentioned in section \ref{sec:coupled wire}, indeed the construction of STO (\ref{biqh:interwire}) is similar to a pure 2d $Z_2$ gauge theory (\ref{zk gauge theory:interwire}), except that the microscopic building blocks are replaced from 1d spin chains (\ref{lag:zk gauge theory}) for usual $Z_2$ gauge theory to the 2d $U(1)$-SPT edge states (\ref{biqh:edge wires}). This difference leads to the anomaly of the e$C$m$C$ STO here.

After identifying the anomalous STO of 3d GSPT phase $\gspt{\text{BIQH}}$, we have demonstrated that $\gspt{\text{BIQH}}$ is a nontrivial 3d GSPT phase with on-site $U(1)$ symmetry. Therefore we established the $\mbz_2\times\mbz_2$ classification of 3d GSPT phase with on-site symmetry group $G_0=U(1)$, as summarized in TABLE \ref{tab:a}. These are bosonic analogs of glide-protected topological insulators\cite{Fang2015,Shiozaki2016}, generated by 2 root phases $\gspt{E_8}$ and $\gspt{\text{BIQH}}$, characterized by anomalous STOs e$f$m$f$ and e$C$m$C$ respectively.\\

\subsection{$U(1)$ charge and time reversal symmetry: $\mathbb{Z}_{2}$ classification}\label{sec:BGUT}

With both $U(1)$ charge and $\mbz_2^\bst$ time reversal symmetries, the 2d SRE phases are classified by $H^3\big(U(1)\rtimes Z_2^\bst,U(1)\big)=\mbz_2$. Note that $E_8$ states with chiral edge modes necessarily break time reversal and hence are forbidden here. The 2d root state for this $\mbz_2$ classification is the bosonic quantum spin Hall (BQSH) state\cite{Lu2012a,Levin2012} $\dket{\text{BQSH}}$. Its protected edge states are still described by Lagrangian (\ref{edge:biqh}), with the same ${\bf K}$ matrix but a different charge vector
\bea
{\bf K}=\bpm0&1\\1&0\epm,~~~{\bf q}_{\text{BQSH}}=\bpm1\\0\epm
\eea
The edge chiral bosons $\{\phi^{1,2}\}$ transform under symmetries as
\bea\label{sym:U(1):bqsh}
&\bpm\phi^1\\ \phi^2\epm\overset{e^{\imth\alpha\hat Q}}\longrightarrow \bpm\phi^1\\ \phi^2+\alpha\epm\\
&\label{sym:trs:bqsh}\bpm\phi^1\\ \phi^2\epm\overset{\bst}\longrightarrow \bpm\phi^1+\pi\\-\phi^2\epm
\eea

From general arguments in section \ref{sec:general strategy}, the 3d GSPT phases with on-site $G_0=U(1)\rtimes Z_2^\bst$ symmetry have at most a $\mbz_2$ classification from its 2d root state $\dket{\text{BQSH}}$. The question is, is the 3d state $\gspt{\text{BQSH}}$ out of the coupled layer construction a nontrivial GSPT or not? The answer is yes. Below we construct and analyze the anomalous STO on [001] side surface of $\gspt{\text{BQSH}}$ state, therefore establishing the $\mbz_2$ classification of 3d GSPT phases with onsite $U(1)\rtimes\bst$ symmetry.

We construct the gapped symmetric STO by depositing a 1d quantum wire, \ie 1d spin chains described by (\ref{commutator:spin chain})-(\ref{chiral boson:spin chain}), between every two neighboring layers (see FIG. \ref{fig:glide4}), and coupling all gapless modes on the side surface in proper ways. In particular as illustrated in FIG. \ref{fig:glide4}, we label these bosonic modes in each spin chain by $(\varphi_n,\theta_n)$ with Lagrangian density
\bea\label{bqsh:deposited wires}
\mathcal{L}_1=\frac{1}{2\pi}\sum_{n\in\mbz/2}\partial_t\varphi_n\partial_y\theta_n+\cdots
\eea
and commutation relation
\bea\label{commutator:deposited wire}
&[\varphi_m(x),\theta_n(y)]=[\theta_n(x),\varphi_m(y)]=\pi\imth\delta_{m,n}\text{Sign}(x-y).~~~
\eea
After depositing the spin chains between 2 neighboring layers, the whole surface states are described by
\bea\label{lag:bqsh:sto}
\mathcal{L}_{STO}=\mathcal{L}_0+\mathcal{L}_1+\mathcal{H}_{int}
\eea
where $\mathcal{L}_0$ given in (\ref{biqh:edge wires}) describes the edge states of BQSH layers. Under $U(1)$ and time reversal symmetries, chiral boson fields of the BQSH edge states $\phi_n^{1,2}$ and deposited spin chains $(\varphi_n,\theta_n)$ transform as
\bea\label{sym:U(1)+trs:deposited wire}
\bpm\phi^1_n\\ \phi^2_n\\ \varphi_n\\ \theta^n\epm\overset{e^{\imth\alpha\hat Q}}\longrightarrow\bpm\phi^1_n\\ \phi^2_n+\alpha\\ \varphi_n\\ \theta^n\epm,~~~
\bpm\phi^1_n\\ \phi^2_n\\ \varphi_n\\ \theta^n\epm\overset{\bst}\longrightarrow\bpm\phi^1_n+\pi\\-\phi^2_n\\ \varphi_n+\pi\\-\theta^n\epm.
\eea
Under glide symmetry operation, these chiral boson fields transform as
\bea\label{sym:glide:bqsh}
\bpm\phi^1_{n}\\ \phi^2_n\\ \varphi_n\\ \theta_n\epm\overset{\mcg_y}\longrightarrow\bpm\phi^1_{n+\frac12}\\ \phi^2_{n+\frac12}\\ \varphi_{n+\frac12}\\-\theta_{n+\frac12}\epm
\eea

As shown in FIG. \ref{fig:glide4}, we consider the following interwire terms that couple 3 neighboring 1d Luttinger liquids together:
\bea\label{bqsh:interwire}
&\mathcal{H}_{int}=\sum_{n\in\mbz/2}C_0\cos\hat L_n^0+C_1\cos\hat L_n^1,\\
&\notag\hat L_n^0=\varphi_n-k\phi^1_n+\varphi_{n+\frac12},\\
&\notag\hat L_n^1=\phi_n^2-k(-1)^{2n+1}\theta_{n+\frac12}-\phi^2_{n+\frac12}.
\eea
Clearly the above interwire couplings preserve glide and all onsite symmetries, if we choose the integer $k=$even.

The inter-wire couplings (\ref{bqsh:interwire}) again stablize a toric-code-type $Z_k$ gauge theory on the GSPT surface. The anyons (generated by gauge charge $e$ and gauge flux $m$) in this STO are given by
\bea
&e\sim e^{\imth\phi^e_n},~~~\phi^e_n=\frac1k\varphi_n,\\
&m\sim e^{\imth\phi^m_n},~~~\phi^m_n=-\theta_n-\frac1k\phi_n^2,~~~n\in\mbz.
\eea
These anyons obey $Z_k\times Z_k$ fusion rules
\bea
e^k\sim m^k\sim1.
\eea
as in a 2d $Z_k$ gauge theory described by ${\bf K}_{Z_k}=-\bpm0&k\\k&0\epm$. How do these anyons transform under various symmetries? First of all, $m\sim e^{\imth\phi^m_n}$ carries fractional ($-1/k$) charge of the $U(1)$ symmetry while $e\sim e^{\imth\phi^e_n}$ is charge neutral:
\bea
\bpm\phi^e_n\\ \phi^m_n\epm\overset{e^{\imth\alpha\hat Q}}\longrightarrow \bpm\phi^e_n\\ \phi^m_n-\frac\alpha k\epm
\eea
While $m$ is invariant under time reversal symmetry $\bst$, $e$ transforms as a ``Kramers doublet'' of $\bst$ (for $k$=~even):
\bea
\bpm\phi^e_n\\ \phi^m_n\epm\overset{\bst}\longrightarrow \bpm\phi^e_n+\frac\pi k\\-\phi^m_n\epm\Longrightarrow \bst^k e^{\imth\phi^e_n}\bst^{-k}=-e^{\imth\phi^e_n}.
\eea
These symmetry operations can all be realized in a pure 2d system, \eg by gauging fermion parity in a quantum spin Hall insulator\cite{Ran2008a,Lu2014c} for simplest $k=2$ case. The anomaly of this STO lies in its glide symmetry implementation\cite{Lee2016}, which we reveal below.

Since glide operation changes spatial locations, we need to be able to write down anyon operators associated with one certain anyon type at different spatial locations. As discussed in Appendix \ref{sec:anyon}, this is determined by anyon hopping operators: two anyons of the same type can hop into each other, via a string of local operators invariant under any global symmetry. In particular the string takes the form of $\hat O^a_{n_1,n_2}=e^{\imth\hat T^a_{n_1,n_2}}$, where $a$ labels the anyon type and
\bea\label{def:hopping operator}
T^a_{n_1,n_2}=\phi^a_{n_2}-\phi^a_{n_1}+\sum_{I,n}t_n^I L_n^I,~~~t_n^I\in\mathbb{R}.
\eea
Here $\hat L_n^I$ are the arguments of cosine terms in inter-wire tunneling terms (\ref{biqh:interwire}), and $e^{\imth\phi^a_n}$ creates an anyon $a$ at location $n$. In our STO case, knowing the anyon operators (\ref{anyon:eCmC}) at location $n$, we can use hopping operators to identify anyon operator at location $n+\frac12$. It's straightforward to work out their hopping operators (\ref{def:hopping operator}):
\bea
&T^e_{n,n+\frac12}=0,~~t_n^0=\frac1k\Longrightarrow\phi^e_{n+\frac12}=\phi_n^1-\frac1k\varphi_{n+\frac12},~~\\
&T^m_{n,n+\frac12}=\theta_n,~~t_n^1=-\frac1k\Longrightarrow\phi^m_{n+\frac12}=\theta_{n+\frac12}-\frac1k\phi^2_{n+\frac12}.~~
\eea



%
%
It's straightforward to see the anyon hopping operators $e^{\imth T^{e/m}_{n,n+\frac12}}$ remain invariant under $U(1)$ charge and time reversal symmetries. Notice that for a fixed anyon operator $\phi^a_{n}$ on link $n$, one can always redefine the anyon hopping operator $T^a_{n,n+\frac12}$ and anyon $\phi^a_{n+\frac12}$ on link $n+\frac12$ as
\bea
&\phi^a_{n+\frac12}\rightarrow\phi^a_{n+\frac12}+2M_a\phi^1_n,~~~M_a\in\mbz,\\
&T^a_{n,n+\frac12}\rightarrow T^a_{n,n+\frac12}+2M_a\phi^1_n.\notag
\eea
so that all symmetry requirements for anyon hopping operators remain satisfied.

Therefore under a generic glide symmetry operation, the anyons transform as
\bea\notag
&e^{\imth\phi^e_n}\overset{\mcg_y}\longrightarrow e^{\imth(\frac1k\varphi_{n+\frac12}+2M_e\phi^1_n)}=e^{\imth(2M_e+1)\phi_n^1}e^{-\imth\phi^e_{n+\frac12}},\\
&e^{\imth\phi^m_n}\overset{\mcg_y}\longrightarrow e^{\imth2M_m\phi^1_n}e^{\imth\phi^m_{n+\frac12}}.
\eea
While $m$ anyon can simply be relocated spatially under glide operation, $e$ anyon is attached to a local boson $e^{\imth\phi_n^1}$ after glide operation. More crucially this local boson transforms nontrivially under time reversal operation $\bst$
\bea
b_n\equiv e^{\imth\phi_n^1}\overset{\bst}\longrightarrow e^{-\imth(\phi_n^1+\pi)}=-b_n^\dagger.
\eea
To fully describe the Abelian STO we must also take this boson into account. Therefore the minimal description of the symmetric STO is a $4\times 4$ ${\bf K}$ matrix and charge vector
    \begin{equation}\label{bqsh:sto}
    \bold{K}_{STO}=\left(\begin{array}{cccc}
    0  &  -k  & 0  & 0\\
    -k  &  0  & 0  & 0\\
    0  &  0  & 0  & 1\\
    0  &  0  & 1  & 0\\
    \end{array}\right),~~
    {\bf q}_{STO}=\left(\begin{array}{c}
    0\\ 1 \\ 1\\ 0
    \end{array}\right),~~k=\text{even}.
    \end{equation}
Under $U(1)$ rotation, the chiral boson fields transform as
\bea
&\notag\vec\phi_n\equiv\bpm\phi^m_n\\ \phi^e_n\\ (-1)^{2n-1}\phi^1_{n-\frac12}\\ \phi^2_{n-\frac12}\epm\overset{e^{\imth\alpha\hat Q}}\longrightarrow
\bpm\phi^m_n-\frac\alpha k\\ \phi^e_n\\ (-1)^{2n-1}\phi^1_{n-\frac12}\\ \phi^2_{n-\frac12}+\alpha\epm\\
&=\vec\phi_n+\alpha{\bf K}^{-1}_{STO}{\bf q}_{STO}.\label{bqsh:sym:u(1)}
\eea
Under time reversal $\bst$ they transform as
\bea
&\notag\vec\phi_n\overset{\bst}\longrightarrow
\bpm-\phi^m_n\\ \phi^e_n+\frac\pi k\\ (-1)^{2n-1}\phi^1_{n-\frac12}+\pi\\-\phi^2_{n-\frac12}\epm={\bf W}_\bst\vec\phi_n+\bpm0\\ \frac\pi k\\ \pi\\0\epm,\\
&{\bf W}_{\bst}=\left(\begin{array}{cccc}
      -1 & 0 & 0 & 0\\
      0 & 1 & 0 & 0\\
      0 & 0 & 1 & 0\\
      0 & 0 & 0 & -1
     \end{array}\right).\label{bqsh:sym:trs}
\eea
Both of these on-site symmetry implementations can be realized in a pure 2d model. However, the glide symmetry implementation in the STO is anomalous:
\bea
&\notag\vec\phi_n\overset{\mcg_y}\longrightarrow
\bpm\phi^m_{n+\frac12}+2 M_m\phi^1_n\\-\phi^e_{n+\frac12}+(2 M_e+1)\phi^1_n\\-(-1)^{2n}\phi^1_n\\\phi^2_n\epm={\bf W}_{\mcg_y}\vec\phi_{n+\frac12},\\
&{\bf W}_{\mcg_y}=\left(\begin{array}{cccc}
      1 & 0 & 2M_m & 0\\
      0 & -1 & 2M_e+1 & 0\\
      0 & 0 & -1 & 0\\
      0 & 0 & 0 & 1
     \end{array}\right),~~~M_{e,m}\in\mbz.\label{bqsh:sym:glide}
\eea

Although the STO (\ref{bqsh:sto}) has zero Hall conductance and zero central charge compatible with glide and time reversal symmetries, the anomaly shows up in a more subtle way. To manifest the anomaly, we gauge a discrete $Z_N$ subgroup of the on-site $U(1)$ symmetry, generated by $\hat R_N\equiv\exp(\imth\frac{2\pi}{N}\hat Q)$. For a pure 2d topological order, gauging any discrete on-site symmetry should lead to a consistent 2d topological order. Therefore inconsistencies in the gauged topological order can serve as a fingerprint of symmetry anomalies in STOs.

Under the discrete $Z_N$ charge rotation, the anyons in STO transform as
\bea
&\notag\vec\phi_n\overset{\hat R_N}\longrightarrow
\vec\phi_n+\frac{2\pi}{N}\bpm-1/k\\0\\0\\1\epm.
\eea
After gauging the discrete symmetry $\hat R_N$, the $Z_N$ fluxes (``symmetry defects'') $\mathcal{F}_{Z_N}$ become dynamical excitations in the gauged topological order. This brings a new type of anyons\cite{Lu2016} into the original topological order (\ref{bqsh:sto})
\bea\label{gauge:bqsh}
\mathcal{F}_{Z_N}\sim e^{\imth\Phi^{Z_N}_n},~~~\Phi^{Z_N}_n=(0,\frac1N,\frac1N,0)\cdot\vec\phi_n.
\eea
In Abelian topological order (\ref{bqsh:sto}), an arbitrary anyon $a$ can be represented as
\bea
a\sim e^{\imth\phi^a_n},~~~\phi^a_n=\vec l_a\cdot\vec\phi_n,~~~\vec l_a\in\mbz^4.
\eea
Geometrically all Abelian anyons can be viewed as living on a $d_K$-dimensional (for $d_K\times d_K$ ${\bf K}$ matrix) integer lattice\cite{Cano2014} $\vec l_a\in\Lambda=\mbz^{d_K}$. The ``primitive Bravais vectors'' of this lattice can be chosen as $\{(e^I)_J=\delta_{I,J}|1\leq I\leq d_K\}$, corresponding to anyons $m$, $e$ and bosons $e^{\imth\phi^{1,2}}$. Since gauging the $Z_N$ symmetry will bring in a new anyon (\ref{gauge:bqsh}), it introduces a new primitive vector
\bea\label{new prim vector:bqsh}
\vec l_{\mathcal{F}_{Z_N}}=(0,\frac1N,\frac1N,0)^T.
\eea
for the anyon lattice of the gauged topological order. After gauging the symmetry, the new topological order has a anyon lattice expanded by all integer vectors and (\ref{new prim vector:bqsh}). The basis of the new anyon lattice $\Lambda_{Z_N}$ can \eg be chosen as
\bea
&\vec l_m=(1,0,0,0)^T,~~~\vec l_e=(0,1,0,0)^T,\notag\\
&\vec l_{\phi^2}=(0,0,0,1)^T,~~~\vec l_{\mathcal{F}_{Z_N}}=(0,\frac1N,\frac1N,0)^T.\label{bqsh:sym:anyon lattice}
\eea
For an arbitrary Abelian anyon labeled by vector $\vec l_a\in\Lambda$, it transforms under time reversal and glide as
\bea
\vec l\overset{\bst}\longrightarrow {\bf W}^T_\bst\vec l,~~~~~\vec l\overset{\mcg_y}\longrightarrow{\bf W}^T_{\mcg_y}\vec l.
\eea
For a symmetric 2d Abelian topological order, its anyon lattice must be invariant under all symmetry operations, \ie any ``lattice site'' (corresponding to one anyon) must be mapped to another lattice site on the same lattice $\Lambda$. In our case of STO (\ref{bqsh:sto}), apparently any integer vectors is mapped to another integer vector by (\ref{bqsh:sym:anyon lattice}). However, the anyon (``$Z_N$ symmetry defect'') labeled by a fractional vector $\vec l_{\mathcal{F}_{Z_N}}$ in the gauged topological order is mapped to a new vector
\bea
\notag&{\bf W}^T_{\mcg_y}\vec l_{\mathcal{F}_{Z_N}}=\big(0,0,\frac{2M_e+1}N,0\big)^T-(0,\frac1N,\frac1N,0)^T,~~M_e\in\mbz,\\
&{\bf W}^T_{\mcg_y}\vec l_{\mathcal{F}_{Z_N}}\notin\Lambda_{Z_N},~~~~\forall~N=0\mod2.\label{bqsh:anomaly:N=even}
\eea
For any odd integer $N=1\mod2$, one can always find an integer $M_e\in\mbz$ so that $\frac{2M_e+1}{N}\in\mbz$ and the new vector ${\bf W}^T_{\mcg_y}\vec l_{\mathcal{F}_{Z_N}}$ still belong to the same anyon lattice. For even integer $N=0\mod2$, on the other hand, glide symmetry maps the original anyon lattice (\ref{bqsh:sym:anyon lattice}) of the gauged topological order into a different lattice\cite{Cho2014} as shown in (\ref{bqsh:anomaly:N=even}). Therefore glide symmetry is broken in the new topological order from gauging $Z_N$ symmetry in STO (\ref{bqsh:sto}). We coin this anomalous STO as e$\mcg_y\bst$m$C$. The anomaly of STO e$\mcg_y\bst$m$C$ on the glide-preserving surface of $\gspt{\text{BQSH}}$ therefore demonstrates that $\gspt{\text{BQSH}}$ is a nontrivial 3d GSPT phase with $U(1)\rtimes Z_2^\bst$ global symmetries.

Physically, unlike PT-breaking thermal/charge Hall response of e$f$m$f$/e$C$m$C$ STOs on the surface of $\gspt{E_8}$ and $\gspt{\text{BIQH}}$ states, here the anomaly of STO e$\mcg_y\bst$m$C$ on the surface of $\gspt{\text{BQSH}}$ state is manifested in the glide symmetry implementations on $\frac{\pi}{N/2}$ flux of $U(1)$ symmetry for any even integer $N$. In the simplest case, $\pi$ flux in the anomalous STO secretly breaks glide symmetry, which it must preserve in any pure 2d system. In a pure 2d Abelian topological order, the orientation-reversing glide symmetry operation ${\bf W}_{\mcg_y}$ must change the sign of ${\bf K}$ matrix\cite{Lu2016,Levin2012a}. On the other hand the anomalous STO (\ref{bqsh:sto}), glide symmetry operation (\ref{bqsh:sym:glide}) changes the form of ${\bf K}$ matrix, while preserving all data (braiding statistics and fusion rules) of the Abelian topological order. Therefore the glide symmetry (\ref{bqsh:sym:glide}) is an ``anyonic symmetry''\cite{Cho2014} that can only be realized in anomalous STOs. Establishing $\gspt{\text{BQSH}}$ as a nontrivial 3d GSPT, we achieve the $\mbz_2$ classification of GSPT phases with onsite $U(1)\rtimes Z_2^\bst$ symmetry, as summarized in TABLE \ref{tab:a}.

\subsection{$U(1)$ spin and time reversal symmetry: $\mathbb{Z}_{1}$ classification}

Unlike the previous case considered in section \ref{sec:BGUT}, $U(1)$ spin rotational symmetry and time reversal corresponds to a different symmetry group $G_0=U(1)\times Z_2^\bst$ and a different classification. In this case, the 2d SRE boson phases are classified by $H^3(U(1)\times Z_2^\bst,U(1))=\mbz_1$, without any nontrivial SPT phases. Therefore the associated 3d GSPT classification is also trivial, as shown in TABLE \ref{tab:a}.

\subsection{Other symmetries}

In this section we consider discrete symmetries, by breaking the $U(1)$ symmetry discussed previously into its $Z_N$ subgroup. The results are summarized in the 3 rows at the bottom of TABLE \ref{tab:a}.

We start with on-site symmetry group $G_0=Z_N$, generated by discrete $Z_N$ rotation $\hat R_N$ satisfying $(\hat R_N)^N=1$. The situation is quite similar to $G_0=U(1)$ case discussed in section \ref{sec:BGU}. The 2d SRE phases have a $\mbz_N\times\mbz$ classification. The root phase associated with the integer index $\nu\in\mbz$ is again the chiral $E_8$ state $\dket{E_8}$, leading to 3d GSPT phase $\gspt{E_8}$ featured by anomalous STO e$f$m$f$ (see section \ref{sec:BG}). The other root state for cyclic $q\in\mbz_N$ index is the $Z_N$-SPT phase $\dket{q=1\mod N}$, whose edge states (\ref{edge:biqh}) transform under $Z_N$ rotation as
\bea
\bpm\phi^1\\ \phi^2\epm\overset{\hat R_N}\longrightarrow \bpm\phi^1+\frac{2\pi}N\\ \phi^2+\frac{2\pi}Nq\epm
\eea
These 2d $Z_N$ SPT phases obey the following $\mbz_N$ addition rule
\bea
\dket{q_1}\oplus\dket{q_2}=\dket{q_1+q_2\mod N}.
\eea
Therefore for any odd integer $N$, any $Z_N$-SPT phase can be viewed as the sum of two identical $Z_N$-SPT phases:
\bea
\dket{q}=\dket{\frac{N+q}2}\oplus\dket{\frac{N+q}2},~~~&q=\text{odd},\\
\notag \dket{q}=\dket{\frac{q}2}\oplus\dket{\frac{q}2},~~~&q=\text{even}.
\eea
As a result according to $Z_2$ addition rule (\ref{z2 addition rule}) of 3d GSPT phases, none of these 2d $Z_N$-SPT phases will lead to a nontrivial 3d GSPT phase if $N$=odd. On the other hand, this argument stops working for the root $Z_N$-SPT phase $\dket{q=1\mod N}$ when $N$=even. In this case, the surface states (\ref{biqh:edge wires}) on [001] side surface can be symmetrically gapped out by same interwire couplings (\ref{biqh:interwire}) as the $G_0=U(1)$ case. The resulting STO is again toric-code-type $Z_2$ topological order e$C$m$C$, where both $e$ and $m$ anyons carry projective representation of $Z_N$ symmetry. Unlike in $G_0=U(1)$ case, the Hall conductance of symmetric STO is not well defined here due to lack of continuous $U(1)$ symmetry.

To diagnose the STO anomaly for $N$=even case, we gauge the discrete $Z_N$ symmetry generated by $\hat R_N$. In particular, discrete $\hat R_N$ flux (``symmetry defects'')
\bea
\mathcal{F}_{Z_N}\sim e^{\imth(\phi^e-\phi^m)/N}
\eea
become new emergent anyons in the gauged topological order. Among them, the $\pi$ flux
\bea\label{pi flux:Zn STO}
\mathcal{F}_{\pi}\sim e^{\imth\frac{\phi^e-\phi^m}2}
\eea
corresponds to a bound state of $N/2$ elementary $\hat R_N$ fluxes and preserves glide symmetry. The statistical angle of $\pi$ flux (\ref{pi flux:Zn STO}) is $\theta_{\mathcal{F}_{\pi}}=\pi/4\mod\pi$. Since this new anyon $\mathcal{F}_\pi$ corresponding to $\pi$ flux is invariant under glide operation which reverses the statistical angle, we must have
\bea
\mathcal{F}_{\pi}\overset{\mcg_y}\longrightarrow\mathcal{F}_{\pi}\Longrightarrow \theta_{\mathcal{F}_{\pi}}=-\theta_{\mathcal{F}_{\pi}}\mod2\pi.
\eea
Hence the self statistics $\theta_{\mathcal{F}_{\pi}}=\pi/4\mod\pi$ and the gauged STO would violate the orientation-reversing glide symmetry in a pure 2d system. This dictates the anomalous glide symmetry implementation in the $Z_N$-symmetric STO of $N$=even GSPT phase $\gspt{q=1\mod N}$. Therefore we have shown that 3d GSPT phase $\gspt{q=1\mod N}$ is nontrivial only when $N$=even, hence arriving at the $\mbz_{(N,2)}\times\mbz_2$ classification of $G_0=Z_N$ 3d GSPT phases.\\

Next we discuss $G_0=Z_N\rtimes Z_2^\bst$ case with discrete $Z_N$ charge conservation (generated by $\hat R_N$) and time reversal, as summarized in the 2nd row from the bottom of TABLE \ref{tab:a}. The 2d SRE phases have a $\mbz_2\times\mbz_2$ classification for integer $N=$~even, or a trivial classification for $N=$~odd\cite{Lu2012a}. For nontrivial $N=$~even case, there are 2 SRE root phases with gapless edge modes protected by $Z_N\rtimes Z_2^\bst$ symmetry. One is an analog of BQSH state with discrete $Z_N$ symmetry, coined as $\dket{Z_N\text{-BQSH}}$, whose edge states (\ref{edge:biqh}) transform as
\bea\label{edge:zN bqsh}
&\bpm\phi^1\\ \phi^2\epm\overset{\hat R_N}\longrightarrow \bpm\phi^1\\ \phi^2+\frac{2\pi}N\epm,~~~\bpm\phi^1\\ \phi^2\epm\overset{\bst}\longrightarrow \bpm\phi^1+\pi\\-\phi^2\epm.
\eea
These edge states need the protection of both $\hat R_N$ and time reversal symmetry. In comparison, the other root phase is closer to BIQH state with discrete $Z_N$ symmetry, whose edge states (\ref{edge:biqh}) transform as
\bea\label{edge:zN biqh}
&\bpm\phi^1\\ \phi^2\epm\overset{\hat R_N}\longrightarrow \bpm\phi^1+\pi\\ \phi^2+\frac{2\pi}N\epm,~~~\bpm\phi^1\\ \phi^2\epm\overset{\bst}\longrightarrow \bpm\phi^1\\-\phi^2\epm.
\eea
Unlike the previous case, these gapless edge modes remain stable even if time reversal is broken. In both cases the construction of STO is exactly the same as (\ref{bqsh:interwire}) for $\gspt{\text{BQSH}}$ state, which lead to a $Z_k$ ($k=$~even) gauge theory on the surface. The anomaly of their STOs again lies in the glide action on $\pi$ flux (\ie bound state of $N/2$ symmetry defects $\mathcal{F}_{Z_N}$) of the STO: the $\pi$ flux secretly breaks glide symmetry in the sense that anyon lattice is not invariant under glide operation after gauging $Z_N$ symmetry. The first case (\ref{edge:zN bqsh}) is exactly the same as e$\mcg_y\bst$m$C$ STO in GSPT state $\gspt{Z_N\text{-BQSH}}$. For the latter case (\ref{edge:zN biqh}), the anyons $\vec\phi_n\equiv(\phi^m_n,\phi^e_n,(-1)^{2n-1}\phi^1_{n-\frac12},\phi^2_{n-\frac12})^T$  in the anomalous STO transform as
\bea
\vec\phi_n\overset{\hat R_N}\longrightarrow\vec\phi_n+\frac{2\pi}{N}\bpm-\frac1k\\0\\ \frac{N}2\\1\epm,~~~\vec\phi_n\overset{\bst}\longrightarrow{\bf W}_\bst\vec\phi_n.
\eea
where matrix ${\bf W}_\bst$ is defined in (\ref{bqsh:sym:trs}). The glide symmetry operation is the same as in (\ref{bqsh:sym:glide}). Again gauging $\hat R_N$ symmetry leads to new anyon (symmetry defect):
\bea
\mathcal{F}_{Z_N}=(0,\frac1N,\frac1N,\frac12)\cdot\vec\phi_n
\eea
and it's straightforward to verify that the anyon lattice of gauged topological order is not invariant under glide operation (\ref{bqsh:sym:glide}). Since after glide operation each $e$ particle is dressed with a boson $e^{\imth\phi_n^1}$ that transforms nontrivially under $Z_N$ symmetry, we coin this anomalous STO as e$\mcg_yC$m$C$. Therefore we establish the $\mbz_2\times\mbz_2$ classification and two root phases of 3d GSPT phases with $G_0=Z_n\rtimes Z_2^\bst$, as summarized in TABLE \ref{tab:a}.\\

Finally we discuss $G_0=Z_N\times Z_2^\bst$ case, with discrete $Z_N$ spin rotation $\hat R_N$ and time reversal symmetry. Again the 2d SRE phases have a $\mbz_2\times\mbz_2$ classification if $N=$~even, or a trivial one if $N=$~odd. When $N=$~even, one root SPT phase has edge states (\ref{edge:biqh}) that transform under symmetries as
\bea\label{zN x trs:1}
&\bpm\phi^1\\ \phi^2\epm\overset{\hat R_N}\longrightarrow \bpm\phi^1\\ \phi^2+\pi\epm,~~~\bpm\phi^1\\ \phi^2\epm\overset{\bst}\longrightarrow \bpm\phi^1+\pi\\-\phi^2\epm.
\eea
while the other root phase has
\bea\label{zN x trs:2}
&\bpm\phi^1\\ \phi^2\epm\overset{\hat R_N}\longrightarrow \bpm\phi^1+\frac{2\pi}{N}\\ \phi^2+\pi\epm,~~~\bpm\phi^1\\ \phi^2\epm\overset{\bst}\longrightarrow \bpm\phi^1\\-\phi^2\epm.
\eea
In their corresponding 3d GSPT phases, the anomalous STOs are completely similar to those previously discussed in $G_0=Z_N\rtimes Z_2^\bst$ case. In a coupled wire construction in parallel to (\ref{bqsh:interwire}), we can obtain the anomalous STOs: e$\mcg_y\bst$m$C$ for (\ref{zN x trs:1}) case and e$\mcg_yC$m$C$ for (\ref{zN x trs:2}) case. Therefore we establish the $\mbz_2\times\mbz_2$ classification for 3d GSPT phases with $G_0=Z_N\times Z_b^\bst$ onsite symmetry when $N=$~even, as summarized in the bottom row of TABLE \ref{tab:a}.

\begin{table*}
\centering
\begin{tabular}{ | c |c|| c | c || c |c |}
    \hline
    AZ class&\multirow{2}{2.7cm}{On-site symmetry group $G_0$}&\multirow{2}{2.5cm}{Classification of 2d SRE phases}& 2d SRE root phases &\multirow{2}{2.5cm}{Classification of 3d GSPT phases}&Anomaly of STOs\\
    &&&&&\\ \hline
    D& $Z_2^{\bold{P}_f}$& $\mbz$ & $p_x+\imth p_y$ & $\mbz_2$&$c_-=1/4\mod 1/2$ \\ \hline
     BDI&$Z_2^{{\bold P}_f}\times Z_2^{\bst},~\bst^2=1$ & $\mathbb{Z}_1$ & None & $\mathbb{Z}_{1}$& None\\ \hline
    DIII&$Z_4^{\bst},~\bst^2=\bold{P}_f$ & $\mathbb{Z}_{2}$ & $(p_x+\imth p_y)_\uparrow\otimes(p_x-\imth p_y)_\downarrow$  & $\mathbb{Z}_{2}$& $c_{-,\uparrow/\downarrow}=\pm\frac14\mod\frac12$ \\ \hline
       A&$U(1)$ & $\mathbb{Z}\times\mathbb{Z}$ & IQH/neutral~$E_{8}$ &$\mathbb{Z}_{2}\times\mathbb{Z}_{2}$&T-Pfaffian/e$f$m$f$ \\ \hline
             AI &$U(1)\rtimes Z_2^{\bst},~\bst^2=1$ & $\mathbb{Z}_{2}$& BQSH of Cooper pairs & $\mathbb{Z}_{2}$& e$\mcg_y\bst$m$C$ ($Z_2$ gauge theory)\\
    \hline
    AII&$U(1)\rtimes Z_4^{\bst},~\bst^2=\bold{P}_{f}$ & $\mathbb{Z}_{2}$& QSH & $\mathbb{Z}_{2}$& e$\mcg_y\bst$m$C$ ($Z_4$ gauge theory)\\ \hline
    AIII&$U(1)\times Z_2^{\bst}$ & $\mathbb{Z}_{1}$& None & $\mathbb{Z}_{1}$&None\\ \hline
\end{tabular}
\caption{Summary of 3d fermion GSPT phases with various on-site symmetries, including $U(1)$ charge/spin conservation and time reversal ($\bst$) symmetries. The nontrivial GSPT phase in class AII hosts the ``hourglass fermion'' surface states\cite{Wang2016b} in the weak-interaction limit.}
\label{tab:b}
\end{table*}

\section{Fermionic GSPT phases}\label{sec:F}

Since the coupled layer construction (\ref{fixed point wf:2d label}) applies to both boson and fermion systems, we follow the same strategy to classify 3d GSPT phases of interacting fermions. Since 2d SRE phases of interacting fermions have been understood well\cite{Chiu2016}, we can start from these 2d SRE root phases and further constrain the 3d fermion GSPT classification by $Z_2$ addition rule (\ref{z2 addition rule}). Finally we establish a nontrivial 3d GSPT phase by constructing its symmetry-preserving STO and analyzing the anomaly of its STO.

The on-site symmetry for fermions is more subtle than the bosonic case, for it is actually a central extension of the symmetry group where the center is the fermion parity ${\bold P}_f$. Like in the boson case we will focus on $U(1)$ charge/spin and time reversal symmetries, and the associated fermion symmetry can be labeled by Altland-Zirbauer's 10-fold way of symmetry classes, as the 6 classes summarized in TABLE \ref{tab:b}. We focus on these 6 symmetry classes, while our approach can be easily applied to any on-site symmetry.

Previously, the surface topological orders of anti-ferromagnetic (AFM) topological insulators (class A) and superconductors (class D) have been constructed explicitly in Ref. \cite{Mross2015} and \cite{Sahoo2016}. These AFM topological phases preserves a combination $\tilde T_x\equiv T_x\cdot\bst$ of lattice translation $T_x$ and time reversal. As has been discussed in section \ref{sec:loc unitary}, the AFM-SPT phases have the same fixed-point wavefunction (\ref{fixed point wf:2d label}) as GSPT phases, and hence the same coupled layer construction and symmetry-preserving STO. The only difference is that glide in GSPT phases is replaced by ``magnetic translation'' $\tilde T_x$ in the AFM-SPT phases. Therefore for symmetry class A and D, we will not write down the explicit construction of STOs but refer interested readers to Ref. \cite{Mross2015,Sahoo2016}.

%
%
%
%

\subsection{Class D: $\mbz_2$ classification} \label{sec:FG}

Any local fermion Hamiltonian always preserves the fermion parity ${\bold P}_f=(-1)^{\hat F}$ as a global (on-site) symmetry, since each term in a local Hamiltonian must contain an even number of fermion creation/annihilation operators. Without any other on-site symmetry, the fermion system belongs to symmetry class D in Altland-Zirnbauer's 10-fold way language\cite{Altland1997}.

The 2d SRE fermion phases in class D have an integer ($\mbz$) classification, labeled by an integer-valued index $\nu\in\mbz$. The $\nu=1$ root phase is the $p_x+\imth p_y$ chiral superconductor $\dket{p_x+\imth p_y}$ of spinless fermions\cite{Read2000}, featured by chiral Majorana edge modes with $c_-=\nu/2=\frac12$. From the coupled layer construction and $Z_2$ addition rule (\ref{z2 addition rule}) of 3d GSPT phases, $\gspt{p_x+\imth p_y}$ is the only possible nontrivial GSPT in symmetry class D (\ie with onsite symmetry $G_0=Z_2^{{\bf P}_f}$). Below we show that $\gspt{p_x+\imth p_y}$ is indeed a nontrivial GSPT hosting anomalous STOs on its glide-invariant surface, therefore establishing $\mbz_2$ classification of fermion GSPT phases in class D.

In the coupled layer construction of $\gspt{p_x+\imth p_y}$ phase, its gapless surface states on glide-invariant [001] side surface are described by
\bea
\mathcal{H}_0=\imth v_F\sum_{n\in\mbz/2}(-1)^{2n}\chi_n\partial_y\chi_n,
\eea
where $\chi_n$ ($\chi_{n+\frac12}$) is the chiral Majorana mode of layer $A_n$ ($B_n$) for $n\in\mbz$, and $v_F$ is the fermi velocity. Now let's deposit 2 quantum wires of spinless free fermion gas between each two neighboring layers ($\chi_n$ and $\chi_{n+\frac12}$) of opposite chirality, described by
\bea
\mathcal{H}_1=\imth v_F\sum_{n\in\mbz/2}\sum_{f=1,2}\sum_{\alpha=\pm}(-1)^{\alpha}\psi^\dagger_{n,f,\alpha}\partial_y\psi_{n,f,\alpha},
\eea
where $f=1,2$ is the ``flavor'' index for deposited quantum wires, and $\alpha=\pm$ is the chirality (``handedness'') index for right/left movers of each quantum wire. These Dirac fermion modes can also be expressed in terms of Majorana modes
\bea
&\chi^1_{n,f,\alpha}\equiv\frac{\psi^\dagger_{n,f,\alpha}+\psi_{n,f,\alpha}}{2},~~\chi^2_{n,f,\alpha}\equiv\frac{\psi^\dagger_{n,f,\alpha}-\psi_{n,f,\alpha}}{2\imth}.
\eea
With the deposited array of quantum wires, one can define the following 9 branches of chiral Majorana modes associated with each layer:
\bea
\Psi_n\equiv(\chi_n,\chi^{1,2}_{n-\frac12,f,(-1)^{2n}},\chi^{1,2}_{n,f,(-1)^{2n}})^T
\eea
and they transform under glide symmetry as
\bea
\Psi_n\overset{\mcg_y}\longrightarrow\Psi_{n+\frac12}.
\eea
This is exactly the same surface states in the anti-ferromagnetic (AFM) topological superconductor (TSC) considered in Ref.\cite{Sahoo2016}, where a symmetric STO is established by certain gapping terms that couple the 9 branches of Majorana wires. The consequent STO is a non-Abelian topological order, whose edge states are described by $SO(3)_3$ Wess-Zumino-Witten (WZW) model with chiral central charge $c_-=9/4$. Such a nonzero chiral central charge is incompatible with orientation-reversing glide operation in a pure 2d topological order, and manifests the anomaly of the glide-symmetric non-Abelian STO. Therefore $\gspt{p_x+\imth p_y}$ is indeed a nontrivial 3d GSPT phase with anomalous STOs, and 3d fermion GSTP phases in AZ symmetry class D is classified by $\mbz_2$ as summarized in TABLE \ref{tab:b}.

\subsection{Class BDI: $\mbz_1$ classification}

Now let's consider time reversal symmetry $\bst$ satisfying $\bst^2=1$, corresponding to on-site symmetry group $G_0=Z_2^\bst\times Z_2^{{\bf P}_f}$ in a fermion system. Such a ``spinless'' time reversal symmetry can be realized in a magnetic superconductor where the combination of time reversal and $\pi$ spin rotation is preserved. In AZ's 10-fold way this corresponds to symmetry class BDI.

2d SRE fermion phases in class BDI has a trivial $\mbz_1$ classification, \ie there is no nontrivial fermion 2d SPT phases. As a result, the 3d GSPT phases in class BDI also has a trivial $\mbz_1$ classification as shown in TABLE \ref{tab:b}.

\subsection{Class DIII: $\mbz_2$ classification}\label{sec:FGT}

Symmetry class DIII corresponds to spin-$1/2$ fermion system with time reversal symmetry $\bst$, satisfying $\bst^2={\bold P}_f$ where fermions transform as Kramers doublets. The on-site symmetry group is hence $G_0=Z_4^{\bst}$ since $\bst^4=1$.
The 2d fermion SRE phases have a $\mbz_2$ classification\cite{Lu2012a}, where the only nontrivial root phase is the helical TSC $\dket{(p_x+\imth p_y)_\uparrow\otimes(p_x-\imth p_y)_\downarrow}$ with a pair of counter-propagating Majorana fermion edge modes. Below we briefly show this 2d root phase leads to a nontrivial 3d GSPT phase in class DIII with anomalous STOs. This establishes the $\mbz_2$ classification of 3d fermion GSPT phases in class DIII.

Clearly the 2d TSC in class DIII can be simply viewed as a tensor product of $\nu=1$ TSC in class D for spin-$\uparrow$ fermions and its time reversal partner. Its helical edge states indeed consist of a spin-$\uparrow$ right-moving Majorana mode and a spin-$\downarrow$ left-moving Majorana mode. Therefore we can simply double the surface degrees of freedom by including both spin-$\uparrow$ and spin-$\downarrow$, and follow the same construction for $\gspt{p_x+\imth p_y}$ state discussed in section \ref{sec:FG}. Naturally the STO of $\gspt{(p_x+\imth p_y)_\uparrow\otimes(p_x-\imth p_y)_\downarrow}$ state is given by $[SO(3)_3]_\uparrow\times[{SO(3)}_{-3}]_{\downarrow}$. This non-chiral STO is obtained by stacking two chiral STOs ($[SO(3)_3]_\uparrow$ and $[{SO(3)}_{-3}]_{\downarrow}$) on top of each other, and time reversal operation transforms one chiral STO (with chiral central charge $c_{-,\uparrow/\downarrow}=\pm9/4$) into the other. Most crucially, each chiral STO (\eg $[{SO(3)}_{-3}]_{\downarrow}$) is invariant under the glide symmetry operation, which is impossible for a pure 2d topological order. Therefore $\gspt{(p_x+\imth p_y)_\uparrow\otimes(p_x-\imth p_y)_\downarrow}$ hosts an anomalous STO, and belongs to a nontrivial GSPT phase in class DIII. This dictates the $\mbz_2$ classification of 3d fermion GSPT phase in class DIII, as summarized in TABLE \ref{tab:b}.

\subsection{Class A: $\mbz_2\times\mbz_2$ classification} \label{sec:FGU}
A continuous $U(1)$ symmetry leads to symmetry class A of fermion systems. In 2d, SRE fermion phases in class A have a $\mbz\times\mbz$ classification, labeled by two integer indices $(\nu,q)$ similar to the 2d boson case. Physically a SRE fermion phase $[\nu,q]$ in class A is characterized by the following chiral central charge (thermal response) and Hall conductance (charge response)
\bea
c_-=8\nu+q,~~~\sigma_{xy}=q\frac{e^2}{h},~~~\nu,q\in\mbz.
\eea
where $e$ is the fundamental charge carried by each fermion. There are 2 SRE root phases: (i) the integer quantum Hall (IQH) state $\dket{\text{IQH}}$ of fermions with $[\nu=0,q=1]$; (ii) the $E_8$ state $\dket{E_8}$ of neutral bosons with $[\nu=1,q=0]$, built from particle-hole excitations of fermions. The 3d GSPT state $\gspt{E_8}$ associated with 2d $E_8$ state of neutral bosonic particle-hole excitations is clearly a nontrivial GSPT phase, which features anomalous $Z_2$ STO ``e$f$m$f$'' as discussed in section \ref{sec:BG}.

Now let's look into 3d GSPT state $\gspt{\text{IQH}}$ constructed by stacking $\sigma_{xy}=e^2/h$ IQH layers of fermions. Its gapless surface states on glide-invariant [001] side surface is described by
\bea
\mathcal{H}_0=\imth v_F\sum_{n\in\mbz/2}(-1)^{2n}\psi^\dagger_{n}\partial_y\psi_n.
\eea
where glide symmetry is implemented as
\bea
\psi_n\overset{\mcg_y}\longrightarrow\psi_{n+\frac12}
\eea
Such an array of staggered 1d chiral fermion wires is exactly the same as the surface states of the 3d AFM-TI, whose symmetric STO is explicitly constructed by proper inter-wire couplings in Ref.\cite{Mross2015,Sahoo2016}. This leads to a non-Abelian STO coined ``T-Pfaffian''\cite{Chen2014a}, which has a chiral central charge $c_-=1/2$ and Hall conductance $\sigma_{xy}=e^2/2h$. Both the thermal and charge Hall responses are contradictory to the glide symmetry in a pure 2d system, demonstrating the anomaly of the glide-symmetric STO. Therefore $\gspt{\text{IQH}}$ is also a nontrivial fermion GSPT phase in class A.

As a result, due to $Z_2$ addition rule (\ref{z2 addition rule}) we achieve the $\mbz_2\times\mbz_2$ classification of 3d fermion GSPT phass in symmetry class A, as summarized in TABLE \ref{tab:b}. The associated root phases are $\gspt{E_8}$ and $\gspt{\text{IQH}}$ states.

\subsection{Class AI: $\mbz_2$ classification}

Class AI corresponds to on-site symmetry group $G_0=U(1)\rtimes Z_2^\bst$ where time reversal symmetry $\bst$ satisfies $\bst^2=1$. Such a ``non-Kramers'' time reversal symmetry can be realized in magnetic insulators, where only the combination of Kramers time reversal and $\pi$ spin rotation is preserved. The 2d SRE fermion phases in class AI have a $\mbz_2$ classification\cite{Lu2012a}, where the nontrivial root phase is the BQSH state $\dket{2e\text{-BQSH}}$ of bosonic charge-$2e$ Cooper pairs. The corresponding 3d GSPT state $\gspt{2e\text{-BQSH}}$ from the coupled layer construction is a nontrivial GSPT phase, characterized by an anomalous STO detailed in section \ref{sec:BGUT}. Specifically, the $\pi$ flux of charge-$2e$ Cooper pairs is not invariant under glide symmetry operation, manifesting the anomaly of the glide-symmetric STO. Note that a $\pi$ flux for charge-$2e$ Cooper pairs is a $\pm\frac\pi2$ flux for charge-$e$ fundamental fermions. Therefore in the STO of $\gspt{2e\text{-BQSH}}$ state, the anomalous glide symmetry operation will not transform a $\pi/2$ flux into a $-\pi/2$ flux, a phenomena impossible in a pure 2d system. In summary, 3d fermion GSPT phases in class AI have a $\mbz_2$ classification, whose root phase is the $\gspt{2e\text{-BQSH}}$ state.

\subsection{Class AII: $\mbz_2$ classification}\label{sec:FGUT}


In a spin-$1/2$ fermion system with time reversal $\bst^2={\bold P}_f$ and $U(1)$ charge symmetries, the on-site symmetry group is $G_0=U(1)\rtimes Z_4^\bst$, corresponding to symmetry class AII. Most topological insulator materials belong to this symmetry class. The 2d fermion SRE phases in class AII have a $\mbz_2$ classification, whose root phase is the 2d quantum spin Hall (QSH) insulator $\dket{\text{QSH}}$ of spin-$1/2$ fermions. Stacking these 2d QSH layers\cite{Ezawa2016} give rise to the ``hourglass fermion'' surface states of the 3d non-symmorphic free-fermion TI\cite{Shiozaki2016,Alexandradinata2016}. Is such a 3d state $\gspt{\text{QSH}}$ a nontrivial GSPT phase in the presence of strong electronic interactiosn? If yes, what are the properties of its anomalous STO? Below we answer these questions by explict construction of the symmetric STO on top of $\gspt{\text{QSH}}$ state.

Since each QSH layer intersects with the glide-symmetric [001] side surface by its gapless helical edge, the gapless [001] surface states of $\gspt{\text{QSH}}$ phase is described by a 2d array of 1d helical edge modes:
\bea
\mathcal{H}_0=-\imth v_F\sum_{n\in\mbz}(\psi^\dagger_{n,R}\partial_y\psi_{n,R}-\psi^\dagger_{n,L}\partial_y\psi_{n,L}),
\eea
Under glide and time reversal operations, the chiral fermion modes transform as
\bea
&\bpm\psi_{n,R}\\ \psi_{n,L}\epm\overset{\mcg_y}\longrightarrow \bpm\psi_{n+\frac12,L}\\-\psi_{n+\frac12,R}\epm,\\
&\bpm\psi_{n,R}\\ \psi_{n,L}\epm\overset{\bst}\longrightarrow \bpm\psi_{n,L}\\-\psi_{n,R}\epm.
\eea
Meanwhile all chiral fermions carry charge-$e$ each. To study interaction effects on the surface states, we bosonize the chiral fermion and obtain the following chiral boson action:
\bea
\mathcal{L}_0=\sum_{n\in\mbz/2}(\partial_t\phi_{n}^R\partial_y\phi_{n}^R-\partial_t\phi_{n}^L\partial_y\phi_{n}^L)+\cdots
\eea
with commutation relations
\bea
[\phi_{m}^{\chi_1}(y_1),\phi_{n}^{\chi_2}(y_2)]=\imth\pi(-1)^{\chi_1}\text{Sign}(y_1-y_2)\delta_{m,n}\delta_{\chi_1,\chi_2}~
\eea
The chiral fermions are related to chiral boson fields by
\bea
\psi_{n,R}\sim e^{\imth\phi_n^R},~~~\psi_{n,L}\sim e^{\imth\phi_n^L}.
\eea
Hence the chiral bosons transform under symmetries as
\bea
&\phi_{n}^{R/L}\overset{e^{\imth\alpha\hat Q}}\longrightarrow\phi_{n}^{R/L}+\alpha,\\
&\bpm\phi_{n}^{R}\\ \phi_{n}^{L}\epm\overset{\bst}\longrightarrow\bpm-\phi_{n}^{L}\\ \pi-\phi_{n}^{R}\epm,\\
&\bpm\phi_{n}^{R}\\ \phi_{n}^{L}\epm\overset{\mcg_y}\longrightarrow\bpm\phi_{n+\frac12}^{L}\\ \phi_{n+\frac12}^{R}+\pi\epm.
\eea
To reveal the connection to $\gspt{\text{BQSH}}$ case studied in section \ref{sec:BGUT}, we reorganize the chiral boson fields $\phi^{R/L}_n$ into a different basis:
\bea
\phi_n^1\equiv(-1)^{2n}\frac{\phi^R_n-\phi_n^L}2,~~~\phi_n^2\equiv{\phi^R_n+\phi_n^L}.
\eea
And it's straightforward to verify their commutation relation (\ref{commutator:biqh}) with ${\bf K}=\bpm0&1\\1&0\epm$, as well as symmetry transformation rules:
\bea
&\bpm\phi_n^1\\ \phi_n^2\epm\overset{e^{\imth\alpha\hat Q}}\longrightarrow \bpm\phi_n^1\\ \phi_n^2+2\alpha\epm,\\
\label{qsh:sym:trs}&\bpm\phi_n^1\\ \phi_n^2\epm\overset{\bst}\longrightarrow \bpm\phi_n^1-(-1)^{2n}\frac\pi2\\ \pi-\phi_n^2\epm,\\
&\bpm\phi_n^1\\ \phi_n^2\epm\overset{\mcg_y}\longrightarrow \bpm\phi_{n+\frac12}^1+\frac\pi2\\ \phi_{n+\frac12}^2+\pi\epm.
\eea
It's straightforward to see that by depositing 1 spin chain (\ref{bqsh:deposited wires})-(\ref{commutator:deposited wire}) between 2 neighboring QSH layers, we can obtain a symmetric STO using the same interwire coupling terms (\ref{bqsh:interwire}) as the $\gspt{\text{BQSH}}$ case. The only difference is, to preserve time reversal symmetry (\ref{qsh:sym:trs}), we must have
\bea
k=0\mod4.
\eea
for interwire couplings (\ref{bqsh:interwire}) and resultant $Z_k$ gauge theory on the side surface. In terms of original chiral boson fields $\{\phi^{R/L}_n\}$, the interwire couplings are written as
\bea\label{qsh:interwire}
&\mathcal{H}_{int}=\sum_{n\in\frac\mbz2}C_0\cos\hat L_n^0+C_1\cos\hat L_n^1,\\
\notag&L_n^0=\varphi_n-(-1)^{2n}\frac k2(\phi^R_n-\phi^L_n)+\varphi_{n+\frac12},\\
\notag&L_n^1=\phi_n^R+\phi_n^L+k(-1)^{2n}\theta_{n+\frac12}-\phi^R_{n+\frac12}-\phi_{n+\frac12}^L.
\eea
Following the same calculations for the $\gspt{\text{BQSH}}$ STO, we can identify the bulk anyons and their symmetry transformations ($\forall~n\in\mbz$)
\bea
&\bpm\phi_n^e=\frac1k\varphi_n\\ \phi_n^m=-\theta_n-\frac{\phi_n^R+\phi_n^L}{k}\epm\overset{e^{\imth\alpha\hat Q}}\longrightarrow\bpm\phi_n^e\\ \phi_n^m-\frac{2\alpha}{k}\epm,\\
&\bpm\phi_n^e\\ \phi_n^m\epm\overset{\bst}\longrightarrow\bpm\phi_n^e+\frac\pi k\\-\phi_n^m-\frac{\pi}{k}\epm.
\eea
Due to non-locality of electron operator $e^{\imth\phi_n^{R/L}}$, here the anyon hopping operators are different from $\gspt{\text{BQSH}}$ case:
\bea
\notag&T^e_{n,n+\frac12}=0,~t_n^0=\frac1k,~t_n^1=-\frac12\Longrightarrow\\
\notag&\phi^e_{n+\frac12}=-\frac{1}k\varphi_{n+\frac12}+\phi_n^R+\frac k2(\theta_{n+\frac12}-\frac{\phi^R_{n+\frac12}+\phi^L_{n+\frac12}}k),~~\\
\notag&T^m_{n,n+\frac12}=\theta_n,~t_n^1=-\frac1k\Longrightarrow
\phi^m_{n+\frac12}=\theta_{n+\frac12}-\frac{\phi^R_{n+\frac12}+\phi^L_{n+\frac12}}k.~~~
\eea
Again notice that we can redefine the anyon $a$ and their hopping operators $T^a_{n_1,n_2}$ by local bosonic operators that transform trivially under all on-site symmetries:
\bea
\phi^a_n\longrightarrow\phi^a_n+2M_a(\phi_n^R-\phi_n^L),~~~M_a\in\mbz.
\eea
From this we can identify the generic glide symmetry operation on anyons:
\bea
&\notag\bpm\phi_n^e\\ \phi_n^m\epm\overset{\mcg_y}\longrightarrow\bpm\frac k2\phi^m_{n+\frac12}-\phi_{n+\frac12}^e\\ \phi_{n+\frac12}^m\epm\\
&+\bpm\phi^R_n\\0\epm+2(\phi_{n}-\phi_{n}^L)\bpm M_e\\M_m\epm.
\eea
Therefore to fully describe the symmetric STO, we need the following $4\times 4$ ${\bf K}$ matrix and charge vector ${\bf q}$
\bea\label{qsh:k mat}
{\bf K}_{STO}=\bpm0&-k&0&0\\-k&0&0&0\\0&0&1&0\\0&0&0&-1\epm,~~~k=0\mod4.
\eea
The associated excitations of the STO transform under global symmetries as
\bea
&\vec\phi_n\equiv\bpm\phi_n^e\\ \phi_n^m\\ \phi_{n-\frac12}^R\\ \phi_{n-\frac12}^L\epm\overset{e^{\imth\alpha\hat Q}}\longrightarrow \bpm\phi_n^e\\ \phi_n^m-\frac{2\alpha}{k}\\ \phi_n^R+\alpha\\ \phi_n^L+\alpha\epm,\\
&\vec\phi_n\overset{\bst}\longrightarrow{\bf W}_\bst\vec\phi_n+\bpm\frac\pi k\\-\frac\pi k\\0\\ \pi\epm,~~{\bf W}_\bst=\bpm1&0&0&0\\0&-1&0&0\\0&0&0&-1\\0&0&-1&0\epm.~~
\eea
Under glide operation, gauge charge $e$ is dressed by an extra electron $\psi_{n,R}\sim e^{\imth\phi^R_n}$:
\bea
&\notag\vec\phi\overset{\mcg_y}\longrightarrow{\bf W}_{\mcg_y}\vec\phi_{n+\frac12}+\bpm0\\0\\0\\ \pi\epm,\\
&{\bf W}_{\mcg_y}=\bpm-1&k/2&1+2M_e&-2M_e\\0&1&2M_m&-2M_m\\0&0&0&1\\0&0&1&0\epm,~M_{e,m}\in\mbz.~~~
\eea
Similar to the bosonic case in section \ref{sec:BGUT}, the above glide operation doesn't preserve the form of ${\bf K}_{STO}$ matrix: it is an ``anyonic symmetry''\cite{Cho2014} that keeps the fractional statistics ($S$ and $T$ matrices) and all bulk data of the STO invariant.

To demonstrate the anomaly of the above glide symmetry operation on STO, we again gauge a discrete $Z_N$ subgroup of the $U(1)$ charge symmetry, generated by $\hat R_N\equiv\exp(\imth\frac{2\pi}{N}\hat Q)$. Under this discrete $Z_N$ charge rotation, the chiral bosons $\vec\phi$ of STO transform as
\bea
\vec\phi\overset{\hat R_N}\longrightarrow
\bpm\phi^e\\ \phi^m-\frac{4\pi}{kN}\\ \phi^R+\frac{2\pi}N\\ \phi^L+\frac{2\pi}N\epm
\eea
After gauging this discrete $Z_N$ symmetry, the $Z_N$ gauge flux
\bea
\mathcal{F}_{Z_N}\sim e^{\imth\Phi_{Z_N}},~~~\Phi_n^{Z_N}=(\frac2N,0,\frac1N,-\frac1N)\cdot\vec\phi_n.
\eea
becomes a new deconfined anyon excitation. Consequently the anyon lattice of the gauged new topological order is expanded by 4 primitive vectors:
\bea
&\notag\vec l_m=(1,0,0,0)^T,~~~\vec l_e=(0,1,0,0)^T,\\
&\vec l_{\psi_R}=(0,0,1,0)^T,~~~\vec l_{\mathcal{F}_{Z_N}}=\frac1N(2,0,1,-1)^T.
\eea
Under time reversal and glide operations, an arbitrary Abelian anyon $a$ labeled by vector $\vec l_a$ is mapped to another vector
\bea
\vec l_a\overset{\bst}\longrightarrow{\bf W}^T_\bst\vec l_a,~~~
\vec l_a\overset{\mcg_y}\longrightarrow{\bf W}^T_{\mcg_y}\vec l_a.
\eea
The anyon lattice is preserved under time reversal, but under glide operation the new vector $\vec l_{\mathcal{F}_{Z_N}}$ is mapped into
\bea
&\notag{\bf W}^T_{\mcg_y}\vec l_{\mathcal{F}_{Z_N}}=\frac1N(-2,k,4M_e+1,1-4M_e)^T\\
&=-\vec l_{\mathcal{F}_{Z_N}}+\frac{(0,k,4M_e+2,-4M_e)^T}N
\eea
Therefore for any even integer $N\geq4$ (equivalent to odd $N\geq3$ due to conserved fermion parity), glide symmetry ${\bf W}_{\mcg_y}$ does not preserve the anyon lattice structure and is anomalous. Physically, this implies that glide symmetry will not transform a $\pi/N$ flux into a $-\pi/N$ flux for $N\geq3$ in the STO, a contradiction for any glide-symmetric topological order in a pure 2d system.

So far, we have shown that the gapless ``hourglass fermion'' surface states on glide-invariant [001] side surface of 3d non-symmorphic TI can be gapped out without breaking any symmetry, through strong electronic interactions described in (\ref{qsh:interwire}). The consequent gapped STO preserves all symmetries, and in particular the glide symmetry is implemented in an anomalous way that is impossible in any pure 2d system. This establishes a bulk-boundary correspondence for fermion GSPT phases in the strong-interacting limit. As a result the 3d fermion GSPT phases in class AII have a $\mbz_2$ classification (see TABLE \ref{tab:b}), where the nontrivial GSPT phase $\gspt{\text{QSH}}$ hosts the hourglass fermion surface states in the weakly-interacting limit.

\subsection{Class AIII: $\mbz_1$ classification}

Finally we consider $U(1)$ spin rotational symmetry and time reversal associated with onsite symmetry group $G_0=U(1)\times Z_2^\bst$, which corresponds to symmetry class AIII. The associated 2d SRE fermion phases have a trivial $\mbz_1$ classification, without any nontrivial bosonic/fermionic SPT phases\cite{Lu2012a}. Therefore the 3d fermion GSPT phases in class AIII also have a trivial $\mbz_1$ classification, see TABLE \ref{tab:b}.

\section{Discussions and Outlook}\label{sec:conclude}

In this paper we discuss a class of ``weak'' SPT phases with symmetry group $\mbz^{\mcg_y}\times G_0$, coined ``GSPT'' phases, protected by glide symmetry $\mcg_y$ and on-site symmetry $G_0$. We show that via a symmetric finite-depth local unitary quantum circuit, any $d$-dimensional GSPT phase can be reduced to a simple fixed-point wavefunction (\ref{fixed point wf:2d label}) described by stacking $(d-1)$-dimensional SRE phases with the same on-site symmetry $G_0$. This not only establishes a coupled layer construction for any GSPT phase, but also allows us to classify $d$-dimensional GSPT phases with the knowledge of $(d-1)$-dimensional SRE phases. Most generally, the classification of SRE phases with onsite symmetry group $G_0$ in $d$ spatial dimension should form an Abelian group, say $\prod_i\mbz_{a_i}$, where $\{a_i\}$ are all integers and we use $\mbz_\infty\equiv\mbz$ to denote the integer group. Due to the $Z_2$ addition rule (\ref{z2 addition rule}) of GSPT phases, the classification of $(d+1)$-dimensional GSPT phases with symmetry group $\mbz^{\mcg_y}\times G_0$ is at most $\prod_i\mbz_{(a_i,2)}$, where $(a,b)$ stands for the greatest common divisor of two integers $a,b$ and we denote $(\infty,2)\equiv2$. For 3d GSPT phases with various $G_0$ studied in TABLE \ref{tab:a}-\ref{tab:b}, the GSPT classifications are all given by $\prod_i\mbz_{(a_i,2)}$. Therefore we conjecture the simple relation:\\

\begin{theorem}
\label{thm:3d GSPT<->2d SRE}
If $d$-dimensional SRE phases preserving on-site symmetry $G_0$ are classified by an Abelian group $\prod_i\mbz_{a_i}$ where $a_i$ are integers and $\mbz_\infty\equiv\mbz$, then $(d+1)$-dimensional GSPT phases preserving symmetry $\mbz^{\mcg_y}\times G_0$ are classified by $\prod_i\mbz_{(a_i,2)}$. Here $(a,b)$ denotes the greatest common divisor of two integers $a,b$ and $(\infty,2)\equiv2$.
\end{theorem}
~\\
To prove this conjecture, one must show that all root states generating the $\prod_i\mbz_{(a_i,2)}$ group corresponds to nontrivial GSPT phases. In all examples worked out in this paper, we establish the anomalous STOs for these root states in the coupled wire construction of their surface states. In other spatial dimensions such as $d=1$ case, STO does not exist. A systematic understanding of interacting topological invariants for GSPT phases can in principle answer this question, which we leave for future works.\\

Our formulation and results can also be applied to crystalline SPT phases protected by other non-symmorphic spatial symmetries, such as anti-ferromagnetic (AFM) topological insulators\cite{Mong2010,Mross2015} preserving a combination $\tilde T_x\equiv \bst\cdot T_x$ of time reversal $\bst$ and translation $T_x$ operations. Clearly all our arguments and construction directly apply to all AFM SPT phases, leading to the same classification and surface anomaly, given the same on-site symmetry group $G_0$. One natural future direction is to apply our formulation to classify and study weak SPT phases protected by other non-symmorphic symmetries, such as screw rotations.\\

In all our examples, we show that strong interactions can symmetrically gap out the glide-invariant [001] side surface, leading to anomalous STOs. In particular in the case of non-symmorphic electron TI (symmetry class AII, section \ref{sec:FGUT}), the ``hourglass fermion'' surface states can be gapped out by proper interactions between itinerant electrons and other spin degrees of freedom (1d spin chains in our coupled wire construction of STO). We notice that recently these non-symmorphic TIs have been proposed to exist in heavy fermion compounds Ce$_3$Bi$_4$Pt$_3$ and CeNiSn\cite{Chang2016}, where both itinerant electrons and local moments play important roles. It's intriguing to see whether the anomalous STOs can emerge out of strong correlations on the surface of these heavy-fermion non-symmorphic TIs.


\acknowledgments

We thank the hospitality of KITP ``topoquant16'' program where this work was finalized, and Aspen Center for Physics where part of this work was performed. This work is supported by startup funds at Ohio State University (FL,BS,YML), in part by the National Science Foundation under Grant No. NSF PHY11-25915 (FL,YML), and in part by by National Science Foundation grant PHY-1066293 (YML).

\bibliographystyle{apsrev4-1_title}
\bibliography{bibs}

\appendix

\section{2d Abelian Topological Orders in the coupled wire construction}\label{sec:anyon}

In this section, we try to systematically address abelian anyon, anyon operator (AO), hopping operator (HO) and loop operator (LO) in coupled wire construction (CWC) frame of 2d topological orders pioneered in Ref.\cite{Kane2002,Teo2014}. These have all been briefly discussed by Ref.\cite{Teo2014} in simple examples, without providing general arguments. We will systematically discuss the definition and uniqueness of AO and HO for Abelian topolocial orders within CWC, clarifying physical observables and gauge redundancy therein. The HO has other names in the literatures, such as open string operator in string-net model\cite{Levin2005}. The HO can be used to distinguish different anyons in the bulk, which will be used to identify anyon transformation rules under glide symmetry operation.


 In the coupled wire construction frame, the inter-wire coupling can be represented as  $H_{int}=\sum_{i} C_{i}\cos(L_{i})$. Assuming all interaction terms $\cos(L_{i})$ are relevant perturbations to Luttinger liquids in decoupled wires. The condition for them to be pinned simultaneously is that $[\cos(L_{i}(x)),\cos(L_{j}(y))]=0$. Assuming negative coefficients $C_{i}<0$, all phase $L_{i}$ will be pinned as integer multiples of $2\pi$ in ground state. Each anyon excitation as a bulk quasiparticle is realized as a soliton/kink of the cosine terms, where one $L_{i}$ is changed by $2\pi$. Explicit quasiparticle operator and its energy can be calculated within the Sine-Golden equations. We call each $L_{i}$ term a ``link'', on which anyon is located. To create a kink of $L_{i}$, quasiparticle operator $e^{\imth Q}$ should be non-commuting with $L_{i}$, namely $[Q(x),L_{i}(y)]=\pi\imth~\text{Sign}(x-y)$. It is easy to check $Q'(x)=Q(x)+\sum_j \alpha_{j}L_{j}(x)$ for any real $\{\alpha_{j}\}$ will have same commutation relation with $L_{i}$ as $Q(x)$, therefore the quasiparticle operator is not unique. Since all $L_{i}$ is charge neutral, the charge for anyon however is uniquely defined. In summary, the anyon and its charge (and other symmetry quantum numbers) are uniquely defined, however, anyon creation/annihilation operators are not unique. Therefore although the anyon excitation is physical, the AO has a ``gauge redundancy''.

Hopping operator is defined as a string of local operators which can create an anyon on one end and annihilate another anyon on the other end. In the framework of coupled wire construction, there are two kinds of hopping operators: one that hops anyons along the wire can be built from local density and current operators\cite{Teo2014}, while the other hops anyons from one link to another. The former has the form of $\hat{O}=e^{\imth(Q(y)-Q(x))}=e^{\imth\int^{y}_{x}\partial_{x'}Q(x')dx'}$, where $\partial_{x'}Q(x')=\bsq\cdot\partial_{x'}\bsphi^{T}(x')$ is a superposition of quasiparticle density operators. We mainly focus on the second kind of hopping operators $\hat{O}_{ij}=e^{\imth T_{ij}}$, which hops an anyon from link-$i$ to link-$j$. The principle for determining the hoping operators relies on the following fact: a hopping operator creates one pair of particle and anti-particle. Below are the central principles to determine the hopping operators: \\
\begin{enumerate}
      \item {\bf Symmetry:} $\hat{O}_{ij}$ must be a string of local operators, which remain invariant under on-site symmetry operations and hence do not carry on-site symmetry quantum numbers. Physically the quantum numbers of a particle and its anti-particle should cancel each other. Therefore $\hat{O}_{ij}$ must be bosonic, charge neutral for $U(1)$ charge conserved system, spinless in spin conserved system and so on.
      \item  {\bf Locality:} Since hopping operator $\hat{O}_{ij}=e^{\imth T_{ij}}$ annihilates one anyon on link $L_{i}$ and creates another anyon on link $L_{j}$, one can write down a general form for the hopping operator as $T_{i,j}=Q_{j}-Q_{i}+\sum_n \alpha_{n} L_{n}$, since $L_{n}$ terms are pinned as constants in the ground state. With a proper choice of real numbers $\{\alpha_n\}$ we can make $\hat O_{ij}=e^{\imth T_{ij}}$ a ``local'' operator constructed out of the microscopic degrees of freedom in the CWC. Specifically, the braiding statistics of an arbitrary anyon with a hopping operator is always trivial.
      \item {\bf Uniqueness:} A hopping operator satisfying the above two requirements is generally not unique, similar to the anyon operator $Q$. In particular operators that hop the same anyon can always differ by a properly-chosen local operator. This provides an equivalent relation between hopping operators: two $T_{ij}$ fields (for hopping operator $\hat{O}_{ij}=e^{\imth T_{ij}}$) that differ by integer multiples of $\{L_{n}\}$ cannot be physically distinguished, and hence belong to the same equivalent class. Therefore we are allowed to make a canonical choice such that $\alpha_{n}\in(-1/2,1/2],~~\forall~n$. In this ``canonical gauge'', a hopping operator is uniquely determined. Similarly the non-unique part of anyon field $Q_i$ can also be fixed. In practice, a finite correlation length in the bulk of a topological order enables us to only consider $L_{n}$ terms between and near link-$i$ and link-$j$ for hoping operator $\hat{O}_{ij}$.

\end{enumerate}

After uniquely determining the hopping operator, a loop operator can also be built from those hopping operators. A loop operator describes creating a anyon and anti-anyon pair, moving one of them around a closed path, and then annihilating them at the original position.

Knowing the anyon hopping operators allow us to elucidate fractional statistics and anyon species in the coupled wire construction of Abelian topological orders. In particular, \emph{two anyons are identical, if and only if there exists a hoping operator defined above that hops one into the other.} One anyon is identical to the conglumerate of multiple other anyons, if there exist a hoping operator that hops this anyon into multiple other anyons. Examples can be found in section \ref{sec:BGU}, \ref{sec:BGUT} and \ref{sec:FGUT}.

\end{document}